\documentclass[%
reprint,
superscriptaddress,
showpacs,
longbibliography,
bibnotes,
amsmath,amssymb,
prb,
floatfix,
]{revtex4-1}

\usepackage{graphicx}
\usepackage{dcolumn}
\usepackage{bm}
\usepackage{multirow}
\usepackage{array}
\usepackage[usenames,dvipsnames]{xcolor}
\definecolor{nblue}{rgb}{0.0, 0.0, 1.0}
\usepackage[colorlinks,linkcolor=nblue,urlcolor=nblue,citecolor=nblue,plainpages=false,pdfpagelabels,breaklinks]{hyperref}
\usepackage{booktabs}
\usepackage{ctable}
\usepackage{upgreek}
\usepackage{mathrsfs}
\usepackage{amssymb}
\usepackage{amsbsy}
\usepackage{color}
\usepackage{cancel}
\usepackage[bbgreekl]{mathbbol}
\usepackage{marginnote}
\usepackage{amsfonts}
\usepackage[caption=false]{subfig}

\newcommand{\lambdazf}{\lambda_{\rm ZF}}
\newcommand{\sigmazf}{\sigma_{\rm ZF}}
\newcommand{\bcen}{\begin{center}}
\newcommand{\ecen}{\end{center}}
\newcommand{\btab}{\begin{tabular}}
\newcommand{\etab}{\end{tabular}}
\newcommand{\bdes}{\begin{description}}
\newcommand{\edes}{\end{description}}

\newcommand{\beq}{\begin{equation}}
\newcommand{\eeq}{\end{equation}}
\newcommand{\bea}{\begin{eqnarray}}
\newcommand{\eea}{\end{eqnarray}}
\newcommand{\non}{\nonumber}

\newcommand{\half}{\frac{1}{2}}
\newcommand{\bary}{\begin{array}}
\newcommand{\eary}{\end{array}}
\newcommand{\benum}{\begin{enumerate}}
\newcommand{\eenum}{\end{enumerate}}
\newcommand{\bitem}{\begin{itemize}}
\newcommand{\eitem}{\end{itemize}}

\newcommand{\bd} { \mbox{\boldmath $d$}}

\newcommand{\bk} { \bm{k} }

\newcommand{\bq} { \bm{q} }

\newcommand{\ket}[1]{| #1 \rangle}

\newcommand{\eqn}[1] {Eqn.~(\ref{#1})}

\newcommand{\sect}[1] {Section~\ref{#1}}

\newcommand{\fig}[1]{Fig.~\ref{#1}}

\newcommand{\tab}[1]{Table~\ref{#1}}

\makeatletter

\newcommand{\Rmnum}[1]{\expandafter\@slowromancap\romannumeral #1@}
\makeatother

\newlength{\myfigwidth}
\newlength{\myhalffigwidth}
\makeatletter
\newcommand{\thickhline}{%
    \noalign {\ifnum 0=`}\fi \hrule height 2pt
    \futurelet \reserved@a \@xhline
}
\newcolumntype{"}{@{\hskip\tabcolsep\vrule width 2pt\hskip\tabcolsep}}

\newcommand{\mylabel}[1]{\label{#1}} 

\newsavebox{\measurebox}

\begin{document}

\title{Recent progress on  superconductors with time-reversal symmetry breaking}
\author{Sudeep Kumar Ghosh}
\email{S.Ghosh@kent.ac.uk}
\affiliation{Physics of Quantum Materials, School of Physical Sciences, University of Kent, Canterbury CT2 7NH, United Kingdom}
\author{Michael Smidman}
\email{msmidman@zju.edu.cn}
\affiliation{Center for Correlated Matter and Department of Physics, Zhejiang University, Hangzhou 310058, China}
\author{Tian Shang}
 \affiliation{Laboratory for Multiscale Materials Experiments, Paul Scherrer Institut, Villigen CH-5232, Switzerland}
\affiliation{Physik-Institut, Universität Z\"{u}rich, Winterthurerstrasse 190, CH-8057 Z\"{u}rich, Switzerland}
\author{James F. Annett}
\affiliation{H. H. Wills Physics Laboratory, University of Bristol, Tyndall Avenue, Bristol BS8 1TL, United Kingdom}
\author{Adrian Hillier}
\affiliation{ISIS Facility, STFC Rutherford Appleton Laboratory, Harwell Science and Innovation Campus, Didcot OX11 0QX, United Kingdom}
\author{Jorge Quintanilla}
\affiliation{Physics of Quantum Materials, School of Physical Sciences, University of Kent, Canterbury CT2 7NH, United Kingdom}
\author{Huiqiu Yuan}
\email{hqyuan@zju.edu.cn}
\affiliation{Center for Correlated Matter and Department of Physics, Zhejiang University, Hangzhou 310058, China}

\date{\today}

\begin{abstract}
Superconductivity and magnetism are antagonistic states of matter. The presence of spontaneous magnetic fields inside the superconducting state is, therefore, an intriguing phenomenon prompting extensive experimental and theoretical research. In this review, we discuss recent experimental discoveries of unconventional superconductors which spontaneously break time-reversal symmetry and theoretical efforts in understanding their properties. We discuss the main experimental probes and give an extensive account of theoretical approaches to understand the order parameter symmetries and the corresponding pairing mechanisms including the importance of multiple bands.

\end{abstract}

\maketitle

\section{Introduction}
Spontaneously broken symmetry is a cornerstone of our current understanding of the physical world~\cite{Anderson393}. The superconducting state is one of the most spectacular examples: by spontaneously breaking global gauge symmetry, a superconductor's wave function becomes macroscopically coherent - electrons in a superconductor (SC) behave analogously to photons in a laser. This review is concerned with unconventional SCs where an additional symmetry is broken in the superconducting state, namely time-reversal symmetry  (TRS). 

In classical physics, time reversal is equivalent to reversing the momenta of all the particles in the system. Time-reversal symmetry refers to the fact that when this is done, the trajectories we obtain are also valid solutions of the equations of motion. Because of this, any thermal average characterizing the macroscopic state of an ergodic system with TRS must be invariant under time-reversal. Since time-reversal flips the sign of angular momenta, and therefore of the magnetic moments, it follows that an ergodic system with TRS cannot have a net magnetization. The same consideration applies to quantum many-body systems though in this case the spin of the particles constitutes an additional contribution to the angular momentum and therefore time-reversal involves the reversal of all spins in addition to the change of sign of all momenta~\cite{Dresselhaus2007}. Mathematically, this is ensured by the following transformation
\beq
\mathcal{T} c^{\dagger}_{\bk\uparrow} = - c_{-\bk\downarrow} \,\,\,\,;\,\,\, \mathcal{T} c^{\dagger}_{\bk\downarrow} = c_{-\bk\uparrow}
\eeq
where $c^{\dagger}_{\bk\sigma}$ is an operator creating an electron with spin $\sigma$ ($\uparrow$, $\downarrow$) and crystal momentum $\bk$ and $\mathcal{T}$ is the TRS operator.

TRS can be broken through the application of external fields or spontaneously. The canonical example of the former in solid-state physics is the electron fluid in a metal in the presence of a magnetic field, leading to Pauli paramagnetism, Landau diamagnetism and the de Haas-van Alphen effect  \cite{Blundell2001}. The most obvious example of the latter is a Stoner ferromagnet, where a self-consistent exchange field leads to a net magnetization. TRS has important consequences in theories of cosmology as well. For example, in the physics of blackholes, the loop quantum gravity theory predicts that the interior of a blackhole must continue into a white hole~\cite{AshtekarPRL2018,AshtekarPRB2018}. This transition of a blackhole to a whitehole is analogous to a bouncing ball in classical physics: a blackhole ``bounces'' and emerges as its time-reversed version, a whitehole.

The superconducting state is a condensate of pairs of electrons, called ``Cooper pairs''. It is characterized by a complex order parameter which is usually a scalar function with an amplitude and a phase characterizing macroscopic quantum coherence. Superconducting properties such as the Meissner  and Josephson effects in so-called ``conventional'' SCs are consequences of spontaneous gauge-symmetry breaking, and are well described by  Bardeen-Cooper-Schrieffer (BCS) theory~\cite{Tinkham2004}. On the other hand, in many so-called ``unconventional'' SCs additional symmetries, such as the space-group symmetries of the crystalline lattice are broken. This review is concerned with SCs that break time-reversal symmetry, which leads to the appearance of spontaneous magnetic fields in the superconducting state-- a particularly dramatic manifestation of unconventional pairing as all known SCs exhibit perfect diamagnetism {\it via} the Meissner effect, supporting a long-held view that magnetism and superconductivity are antagonistic states of matter.

Evidence for TRS breaking was initially detected in a few  highly correlated superconductors using the muon-spin relaxation ($\mu$SR) technique. $\mu$SR has proved ideal  for searching for broken TRS in superconducting systems, since it is a local probe capable of detecting very small magnetic fields in a sample, in the absence of an applied magnetic field. For a number of reasons,  superconductivity with broken TRS has become a topic of particular interest recently. One reason is the discovery of  superconductors with weak-electronic correlations, where signatures of TRS breaking are detected using $\mu$SR, but other superconducting properties appear to largely resemble those of conventional systems. These therefore appear to correspond to a class of materials distinct from the previously known examples of TRS breaking superconductors, and determining the origin of the broken TRS requires further experimental and theoretical attention. Furthermore, even amongst the established examples, various outstanding questions remain. In particular, Sr$_2$RuO$_4$ had long been regarded as the canonical example of a triplet superconductor, with a chiral $p$-wave order parameter \cite{Mackenzie2003}. This understanding has been thrown into question however, by recent nuclear magnetic resonance (NMR) measurements which point to singlet-pairing \cite{Pustgow2019}. This has stimulated a flurry of experimental and theoretical studies aimed at understanding the results. Moreover, in the last decade there has been a growing appreciation of the role-played by topology in determining the properties of condensed matter systems, and topological superconductivity is one of the most sought after goals in this field. As such there has been particular  interest in whether unconventional TRS breaking superconducting states can exhibit this phenomenon.

In general, TRS breaking in the superconducting ground state requires a degenerate instability channel leading to a multi-component order parameter. This is because under the application of the time-reversal operator, the order parameter transforms into another order parameter which is not just a phase multiple of the original one. This is satisfied by an order parameter having multiple components and a nontrivial phase difference between them. The degeneracy of the corresponding instability channel can have many different origins. For example, it can arise from breaking underlying symmetries of the crystal, as in the chiral $p$-wave state~\cite{Mackenzie2003} or the loop-super current state proposed for multi-orbital systems~\cite{Ghosh2018}; or it can arise from breaking the group of spin rotations, as in the nonunitary triplet state with equal spin pairing proposed for LaNiC$_2$ and LaNiGa$_2$~\cite{Hillier2009,Quintanilla2010,Hillier2012,Weng2016,Ghosh2019}.

A desired milestone in the study of unconventional superconductivity is to uniquely determine the structure of the order parameter and the corresponding pairing mechanism. Approaching  this goal for a specific material needs experimental and theoretical knowledge to work in unison. By considering the experimentally observed properties of a system, the possible symmetry-allowed order parameters can be determined. These possibilities are then considered in mean-field theories with a model band structure or full first-principles band structure to predict low-energy properties of the material which are compared with the corresponding experimental data. A crucial means of determining the nature of the pairing state, is the characterization of the magnitude and structure of the superconducting energy gap. This can be probed using a number of techniques, including the measurement of thermodynamic quantities such as the specific heat and magnetic penetration depth, spectroscopic techniques such as angle-resolved photoemission spectroscopy (ARPES) and scanning tunneling spectroscopy, or experiments utilizing Josephson tunneling. The structure and shape of the normal state Fermi surfaces computed from first principles or measured by de Haas-van Alphen and ARPES experiments~\cite{Holstein1973} can also help in narrowing-down the possible order parameter structures. However, very often, mainly due to the high symmetry of the crystal structure of the material in question, the above procedure leads to many different possible order parameters with similar low-energy properties. Thus in practice the unequivocal determination of the structure of the order parameter of a TRS-breaking SC is often very challenging.

This review article is broadly divided into two parts: 1) Recently discovered TRS breaking SCs with discussion of the corresponding experimental techniques and 2) Theoretical understanding of the possible structures of order parameters and low-energy properties of such materials. In \sect{sec:exp-probes} we briefly discuss the main experimental probes used to directly measure and characterize TRS breaking in SCs. \sect{sec:materials} describes systems where TRS breaking has been discovered  in the superconducting state, including  examples in strongly correlated SCs as well as the more recent cases in  materials which generally exhibit fully-open gaps.  A summary of TRS breaking superconductors is also provided in \tab{tab:trsbsc}. We then go on in \sect{sec:GL-theory} to discuss the possible structures of order parameters of SCs based on general symmetry arguments within the Ginzburg-Landau theory, that is without requiring the details of the pairing mechanism and hence should apply to most unconventional SCs. In the next \sect{sec:GL-theory-TRSB}, we focus on the order parameter symmetries of TRS breaking SCs considering some specific examples. In the \sect{sec:MFT} we discuss the method for computing the properties of superconducting ground states within the mean-field approximation, which in its full form requires using a specific model of the pairing interaction and the normal state band-structure. In the following \sect{sec:novel-gs} we discuss recent theoretical proposals of novel superconducting ground states for specific, recently-discovered TRS-breaking superconducting materials. Finally in \sect{sec:summary} we conclude by outlining some possible future directions.

\section{Experimental probes}
\mylabel{sec:exp-probes}
Here, we briefly review the three main experimental probes which can \emph{directly} detect the presence of spontaneous magnetic fields in the superconducting state, so as to detect TRS breaking in SCs: $\mu$SR, the optical Kerr effect and superconducting  quantum interference device (SQUID) magnetometry.  

\subsection{Muon spin  relaxation}
$\mu$SR is a very sensitive local probe of extremely small magnetic fields in a sample (down to $\sim 10^{-5}$T). As a result, $\mu$SR measurements performed in the absence of an external field can reveal the spontaneous appearance of very small magnetic fields in the superconducting state, and hence whether TRS is broken. A particular advantage of this technique is that the presence of TRS breaking can be probed in both single crystal and polycrystalline samples.  More comprehensive descriptions of the $\mu$SR technique can be found in Refs.~\onlinecite{Yaouanc2011,Blundell1999,Sonier2000,Khasanov2015} but here we discuss a few pertinent aspects. 

Spin polarized positive muons are implanted in the sample, which stop at interstitial positions in the crystal lattice corresponding to minima of the Coulomb potential. Muons decay with a half life of 2.2$\mu$s into a positron and two neutrinos, and the emitted positrons are counted via detectors which in the experiments described here are situated in forward (F) and backward (B) positions. The muons' spins precess about the local magnetic fields at the muon stopping sites $\mathbf{B}_{\rm loc}$, at the Larmor frequency  $\omega_\mu = \gamma_\mu |\mathbf{B}_{\rm loc}|$, where $\gamma_\mu/2\pi$ = 135.5~MHz T$^{-1}$ is the muon gyromagnetic ratio. Since the  positrons are emitted preferentially along the direction of the muon spin, the asymmetry $A(t)$ reflects the local field distribution, where

\begin{equation}
A(t)=\frac{N_F(t)-\alpha N_B(t)}{N_F(t)+\alpha N_B(t)}.
\end{equation}

\noindent Here  $N_F(t)$ and $N_B(t)$ are the number of positrons counted at a time $t$ at the forward and backward detectors respectively, while $\alpha$ is a calibration constant. In particular, when there is a distribution of fields in a sample, the muons'  spins precess at different rates, and the broader the field distribution, the more rapidly $A(t)$ decays. Time reversal symmetry breaking is experimentally detected from an increase of the relaxation rate of $A(t)$, arising due to the spontaneous  appearance of additional internal fields in the superconducting state. For a random static Gaussian distribution of fields, $A(t)$ is described by the Kubo-Toyabe relaxation function

\begin{equation}
G_{KT}=\frac{1}{3}+\frac{2}{3}(1-\sigmazf^2t^2)e^{-\frac{\sigmazf^2t^2}{2}},
\end{equation}

\noindent where the Gaussian relaxation rate $\sigmazf$ is proportional to the width of the field distribution. Since nuclear moments are static on the timescale of the muon lifetime, these make a temperature independent contribution to $\sigmazf$. In most of the cases discussed here, the asymmetry was fitted with the product of $G_{KT}$ with a Lorentzian relaxation

\begin{equation}
A(t) = A_0G_{KT}(t)e^{-\lambdazf t},
\label{AsymFit}
\end{equation}

\noindent where $A_0$ is the initial asymmetry and $\lambdazf$ is the Lorentzian relaxation rate. On the other hand, sometimes a combined Lorentzian and Gaussian Kubo-Toyabe function is utilized,

\begin{equation}
A(t) = A_0\frac{1}{3}+\frac{2}{3}(1-\sigmazf^2t^2-\lambdazf t)e^{-\frac{\sigmazf^2t^2}{2}-\lambdazf t}.
\label{AsymFit2}
\end{equation}

\noindent In some experiments, the TRS breaking is manifested by an increase of $\sigmazf$ in the superconducting state, in others it is via $\lambdazf$. Note that $\lambdazf$ is often attributed to magnetic fields  which fluctuate rapidly on the timescale of the muon lifetime, i.e. `electronic' moments, since such a Lorentzian decay results from a Gaussian distribution of fields in the fast fluctuation limit. However in TRS breaking SCs, the increase in $\lambdazf$ or $\sigmazf$ is typically suppressed by a small longitudinal field of around 20~mT. The fact that the muon can be decoupled from its local environment by a small longitudinal field indicates that these internal fields are  quasistatic. This suggests that this extra relaxation does not arise from paramagnetic impurities.

\subsection{Optical Kerr effect}
The optical Kerr effect is another very sensitive probe for detecting spontaneous fields inside a SC by measuring the change in the Kerr angle. Due to the technically challenging nature of these experiments, TRS breaking has been observed using the Kerr effect in only a few SCs. Moreover, the reliable determination of the presence of broken TRS generally requires single crystal, rather than polycrystalline samples. The quantity measured in these experiments is the polar Kerr angle $\theta_{\rm K}$, which corresponds to how much normally incident linearly polarized light is rotated \cite{Xia2006,Schemm2014,Schemm2015,Schemm2017,Levenson2018}. Non-zero $\theta_{\rm K}$ arises from differences between the complex refractive indices of left ($n_{\rm L}$) and right   ($n_{\rm R}$) circularly polarized light via

\begin{equation}
\theta_{\rm K} = -{\rm Im}\left(\frac{n_{\rm R}-n_{\rm L}}{n_{\rm L}n_{\rm R}-1}\right)={\rm Im}\frac{4\pi}{\omega}\left(\frac{1}{n(n^2-1)}\sigma_{xy}\right).
\label{KerrAngleEq}
\end{equation}

\noindent As such, $\theta_{\rm K}$ is related to the presence of an imaginary component to the off-diagonal part of the conductivity tensor $\sigma_{xy}$, which only arises when time reversal symmetry is broken. In SCs this requires that the superconducting order parameter has multiple components with a phase between them \cite{Schemm2017}. In order to detect the TRS breaking signal, it is necessary to  measure a very small $\theta_{\rm K}$ on the order of $1~\mu$rad, yet with a sufficiently low beam intensity to avoid sample heating. These were performed using a Sagnac interferometer, as described in detail in Refs.~\onlinecite{Xia2006,Schemm2014,Xia2006b,Kapitulnik2009} where the authors achieve a sensitivity in $\theta_{\rm K}$  of 10~nrad, at temperatures down to 100~mK. Due to the presence of different domains, upon zero-field cooling (ZFC) there may be no net contribution to $\theta_{\rm K}$. Therefore, in addition to measuring $\theta_{\rm K}$ after cooling in zero-field, a training field was often applied in the superconducting state to align the domains. The field was then turned off, and measurements were performed upon warming to above $T_c$. Upon reversing the direction of the training field, $\theta_{\rm K}$ has the same magnitude but opposite sign below $T_c$, and therefore the magnetic field couples to and reorientates the domains. In the cases of Sr$_2$RuO$_4$, UPt$_3$ and URu$_2$Si$_2$ \cite{Xia2006,Schemm2014,Schemm2015}, the magnitude of  $\theta_{\rm K}$ after applying the training field is the same as in the ZFC measurements, indicating that the domain size is larger than the diameter of the beam ($\approx10.6~\mu$m). 
\begin{table*}[!t]
\centering
\caption{Properties of TRS breaking superconductors. S and N in the space group column denote symmorphic and nonsymmorphic crystal structures respectively, while the column headed TRSB evidence displays the techniques from which TRS breaking was deduced. References not provided in the main text, are also displayed. \label{tab:trsbsc}} 
\begin{ruledtabular}
\begin{tabular}{lcccccccccr}
Compound & TRSB & Structure & Space & Point & Inversion &  Gap structure & Proposed \\
 & evidence & & group & group & &  & state \\
\colrule\\
U$_{1-x}$Th$_x$Be$_{13}$ & $\mu$SR&Cubic&$Fm\bar{3}c$ (S) & O$_h$ & Y & -- & --\\\\
UPt$_3$ & $\mu$SR, Kerr & Hexagonal&$P6_3/mmc$ (N)&D$_{6h}$&Y&line node& $E_{2u}$ triplet\\\\
URu$_2$Si$_2$ & Kerr&Tetragonal&$I4/mmm$ (S)&D$_{4h}$&Y&line + & chiral $d$-wave \\
&&&&&&point nodes&\\\\
UTe$_2$ \cite{Ran2019,Metz2019} &  Kerr&Orthorhombic&$Immm$ (S)&D$_{2h}$&Y&point nodes & non-unitary \\\hline\\
Sr$_2$RuO$_4$ & $\mu$SR, Kerr & Tetragonal & $I4/mmm$ (S) & D$_{4h}$ & Y & line node & chiral singlet \\\hline\\
Pr(Os$_{1-x}$Ru$_x$)$_4$Sb$_{12}$ & $\mu$SR, Kerr & Cubic& $Im\bar{3}$ (S) &T$_h$&Y &full gap, point node& multicomponent  \\
&&&&&&&$d$-wave \\\\
Pr$_{1-y}$La$_y$Os$_4$Sb$_{12} (y<1)$ & $\mu$SR & Cubic& $Im\bar{3}$ (S) &T$_h$&Y &full gap, point node & --\\\\
Pr$_{1-y}$La$_y$Pt$_4$Ge$_{12} (y<1)$ & $\mu$SR & Cubic& $Im\bar{3}$ (S) &T$_h$&Y &point node&non-unitary $p$-wave\\\hline\\
SrPtAs &$\mu$SR&Hexagonal&$P6_3/mmc$ (N) &D$_{6h}$&Y&full gap&chiral $d$-wave \\\hline\\

Re$_{0.82}$Nb$_{0.18}$ & $\mu$SR & Cubic & $I\bar{4}3m$ (S) & T$_d$ & N & full gap & LSC\\
Re$_6$(Zr,Hf,Ti)\\\\
Re & $\mu$SR & Hexagonal & $P6_3/mmc$ (N) & D$_{6h}$ & Y &  full gap &  --\\\hline\\
LaNiC$_2$ & $\mu$SR, SQUID & Orthorhombic & $Amm2$ (S) & C$_{2v}$ & N &  two full gaps & INT \\\\
LaNiGa$_2$ & $\mu$SR & Orthorhombic & $Cmmm$ (S) & D$_{2h}$ & Y & two full gaps & INT \\
\hline\\
La$_7$(Ir,Rh)$_3$ & $\mu$SR & Hexagonal & $P6_3mc$ (N) & C$_{6v}$ & N & full gap & singlet dominated\\
&&&&&&&mixed state\\\hline\\
Zr$_3$Ir \cite{Sajilesh2019,Shang2019} & $\mu$SR & Tetragonal & $I\bar{4}2m$ (S) & D$_{2d}$ & N &  full gap & singlet dominated\\
&&&&&&&mixed state\\\hline\\
(Lu,Y,Sc)$_5$Rh$_6$Sn$_{18}$ & $\mu$SR & Tetragonal & $I4_1/acd$ (N) & D$_{4h}$ & Y &  full gap & multicomponent \\
&&&&&&&singlet or triplet
\end{tabular}
\end{ruledtabular}
\end{table*}

\begin{figure*}[tb]
\begin{center}
\includegraphics[width=\textwidth]{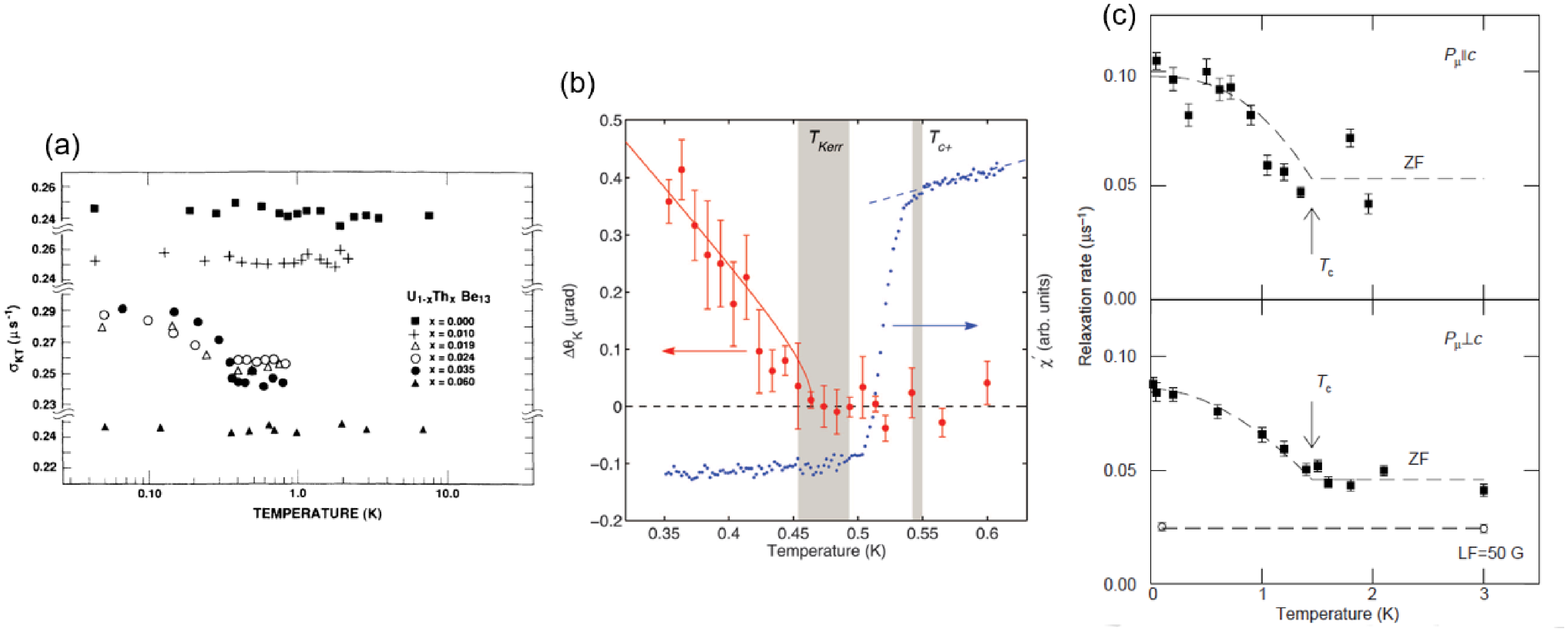}
\end{center}
\caption{(a) Temperature dependence of the Gaussian relaxation rate obtained from zero-field $\mu$SR measurements of U$_{1-x}$Th$_x$Be$_{13}$ for different dopings. For stoichiometric UBe$_{13}$, no change is observed below $T_c$, but for   2-4\% Th-doping, an increase in the relaxation rate is observed below the emergent second superconducting transition. Reproduced with permission from Ref.~\onlinecite{Heffner1990}. Copyright 1990 by the American Physical Society. (b) Temperature dependence of the Kerr angle and mutual inductance of UPt$_3$. The increase of the Kerr angle indicates the breaking of time reversal symmetry, and it can be seen that this onsets at a second transition $T_{\rm Kerr}=0.46$~K, below the superconducting transition at $T_{c+}=0.55$~K.  From Ref.~\onlinecite{Schemm2014}. Reproduced with permission from AAAS. (c) Temperature dependence of the Lorentzian relaxation rate ($\lambdazf$ in Eq.~\ref{AsymFit}) obtained from zero-field $\mu$SR on Sr$_2$RuO$_4$, with the initial muon spin polarized parallel and perpendicular to the $c$-axis. Both quantities show an increase below $T_c$ indicating TRS breaking, while this effect is destroyed by a longitudinal field of 50~mT. Reproduced with permission from Macmillan Publishers Ltd: Nature \cite{Luke1998}, copyright 1998.
 }
\label{CorrTRSBFig}
\end{figure*}

\subsection{SQUID magnetometry}

SQUID magnetometry can in principle be used to detect broken TRS, if the system develops a small net magnetization as it enters the superconducting state. Though the field resulting from such a magnetization will be fully screened by the Meissner effect in the bulk of the sample, in a type-II SC the superconducting order parameter will be suppressed near the surface, allowing the magnetic field due to the bulk magnetization to be detected. One major problem with this approach is that even large domains will be averaged out, since one is measuring the signal from the whole sample. Nevertheless as we discuss below this technique has been successfully employed to detect a net magnetization in LaNiC$_2$~\cite{Sumiyama2015}.

\section{Superconductors with broken time reversal symmetry}
\mylabel{sec:materials}

\subsection{TRS breaking in strongly correlated superconductors}
\label{CorrSec}

The first reported examples of TRS breaking were in a handful of strongly correlated SCs,  the properties of which are included in  \tab{tab:trsbsc}. In such systems, the presence or close proximity of magnetism, together with the large Coulomb repulsion already gave a strong indication that the superconductivity must be unconventional, and is likely mediated by magnetic interactions. UBe$_{13}$ was one of the earliest heavy fermion SCs to be discovered, where it was proposed to be an example of triplet $p$-wave pairing \cite{Ott1984}. For heavy fermion SCs, the extremely large onsite Coulomb repulsion has  been thought to disfavor  $s$-wave superconductivity, and instead pairing states with non-zero orbital angular momentum have generally been anticipated. Unusual behavior was quickly found upon doping with Th in the U$_{1-x}$Th$_x$Be$_{13}$ system, where a second superconducting transition in the specific heat emerges at around $x=0.03$, and disappears at $x=0.06$ \cite{Ott1985}. Zero-field $\mu$SR subsequently revealed that while no change of $\sigmazf$ is found in the superconducting state of stoichiometric UBe$_{13}$, an increase is observed at the second transition induced by Th-doping, as displayed in Fig.~\ref{CorrTRSBFig}(a) \cite{Heffner1990}. One of the scenarios proposed  was that this corresponds to a transition into a state with a complex multicomponent order parameter, which breaks time reversal symmetry \cite{Sigrist1989}.

Perhaps the two canonical examples of SCs with strong electronic correlations exhibiting TRS breaking, although neither without controversy, are Sr$_2$RuO$_4$~\cite{Luke1998,Xia2006,Mackenzie2003} and UPt$_3$~\cite{Luke1993,Schemm2014}. UPt$_3$ is another notable early example of a heavy fermion SC, where two transitions in the zero-field specific heat are observed in stoichiometric samples, clearly indicating an unconventional superconducting order parameter \cite{Fisher1989} (for a detailed review see Ref.~\onlinecite{Joynt2002}). Zero-field $\mu$SR measurements of UPt$_3$ by Luke \textit{et al.} revealed a clear increase of the Lorentzian relaxation rate in the superconducting state below the lower critical temperature \cite{Luke1993}. Here the stoichiometric nature of the system compared to the aforementioned U$_{1-x}$Th$_x$Be$_{13}$  allowed the authors to exclude the scenario of an impurity induced magnetic transition, giving a strong indication of TRS breaking. This result was corroborated much more recently by measurements of the optical Kerr rotation, displayed in Fig.~\ref{CorrTRSBFig}(b) \cite{Schemm2014}. Evidence was also found for TRS breaking in URu$_2$Si$_2$ and very recently, UTe$_2$ using the same technique \cite{Schemm2015,Hayes2020}. In the case of the UTe$_2$, $\mu$SR measurements exhibited a strong temperature dependent relaxation corresponding to ferromagnetic fluctuations, and therefore whether TRS is preserved in the superconducting state could not be probed \cite{Sundar2019}. For Kerr rotation measurements of UPt$_3$, the onset of TRS breaking could be shown to occur at the lower  superconducting transition.  This is expected for the proposal of triplet superconductivity with an $E_{2u}$ order parameter \cite{Sauls1994,Joynt2002}, where the order parameter picks up a second component at the lower transition, giving rise to a complex superposition. One curious aspect is that a subsequent zero-field $\mu$SR study failed to detect an increase of the $\mu$SR relaxation rate within the superconducting state of UPt$_3$ \cite{deReotier1995}. The authors of the later study explain  this discrepancy as being due to a higher sample quality, attributing the previous findings to the presence of defects or impurities.

\begin{table}
\centering
\caption{A list of those strongly correlated superconductors where $\mu$SR or the Kerr effect have been used to provide evidence for either the presence or absence of time-reversal symmetry in the superconducting state. \label{tab:unconv}} 
\begin{tabular}{ccc}
\hline\\
Material & Broken TRS? & Reference \\
\\\hline\\
UPt$_3$ & $\checkmark/\times$&\cite{Luke1993,Schemm2014,deReotier1995}  \\
UBe$_{13}$ & $\times$& \cite{Heffner1990}   \\
U$_{1-x}$Th$_x$Be$_{13}$ & $\checkmark$& \cite{Heffner1990} \\
($0.02\leq x\leq0.04$)  &   \\
URu$_2$Si$_2$ &  $\checkmark$& \cite{Schemm2015}\\
UTe$_2$ & $\checkmark$&\cite{Hayes2020}   \\\hline\\
CeCu$_2$Si$_2$ & $\times$ & \cite{Feyerherm1997}  \\
CeCoIn$_5$ & $\times$ & \cite{Higemoto2002}  \\
CeIrIn$_5$ & $\times$ & \cite{Higemoto2002}  \\\hline\\
YBa$_2$Cu$_3$O$_7$ & $\times$ & \cite{Kiefl1990,Saadaoui2013}\\
Bi$_2$Sr$_{2-x}$La$_x$CuO$_{6+\delta}$ &$\times$& \cite{Kiefl1990,Russo2007}\\\hline\\
Ba$_{1-x}$K$_x$Fe$_2$As$_2$ &$\times$,$\checkmark(x=0.73)$&\cite{Mahyari2014,Grinenko2017}\\\hline\\
Sr$_2$RuO$_4$ & $\checkmark$&\cite{Luke1998,Xia2006}\\\hline\\
Pr(Os$_{1-x}$Ru$_x$)$_4$Sb$_{12}$ & $\checkmark$ & \cite{Aoki2003,Shu2011}\\
Pr$_{1-y}$La$_y$Os$_4$Sb$_{12} (y<1)$ &$\checkmark$ & \cite{Shu2011}\\
Pr$_{1-y}$La$_y$Pt$_4$Ge$_{12} (y<1)$ & $\checkmark$ & \cite{Maisuradze2010,Zhang2019}\\\hline

\end{tabular}
\end{table}

Sr$_2$RuO$_4$ has been one of the most extensively characterized unconventional superconducting systems, owing in a large part to the system having the same structure as the high-$T_c$ cuprates, but it was long thought to be a rare example of spin-triplet superconductivity outside of the U-based heavy fermion SCs (see Ref.~\onlinecite{Mackenzie2003} for a review).  Evidence for TRS breaking was reported from zero-field $\mu$SR measurements on single crystals of Sr$_2$RuO$_4$, as displayed in Fig.~\ref{CorrTRSBFig}(c) \cite{Luke1998}. An increase of the Lorentzian relaxation rate $\lambdazf$ is observed when the muon spin is polarized both parallel and perpendicular to the $c$-axis. Although $\lambdazf$ in $\mu$SR measurements is often associated with rapidly fluctuating fields, the fact that the effect is destroyed by a longitudinal field of 50~mT was taken as evidence that these fields are static on the timescale of the muon lifetime, and they were estimated to have a characteristic strength of 0.5~G.  

The primary evidence for spin-triplet superconductivity came from $^{17}$O NMR measurements, which suggested that the Knight shift, and hence the spin-susceptibility, remains unchanged along all directions upon cooling through the superconducting  transition at $T_c\approx1.5$~K \cite{Ishida1998}. This conclusion was also supported by polarized neutron scattering measurements of the spin susceptibility \cite{Duffy2000}. The combination of TRS breaking and triplet superconductivity naturally led to the proposal of a chiral $p$-wave $p_x+ip_y$ pairing \cite{Mackenzie2003}. However, there continued to be a number of outstanding issues, namely that the Knight shift appeared to be constant along all field directions \cite{Ishida1998}, suggesting that the triplet order parameter $\mathbf{d(k)}$ rotates with the applied field, but the spin-orbit coupling would be expected to be sufficiently strong to pin $\mathbf{d(k)}$ along a particular crystallographic axis. Moreover, $T_c$ did not show the expected cusp in strain experiments \cite{Hicks2014}, and both scanning SQUID and scanning Hall probe  microscopy  measurements did not find evidence for the  currents anticipated to emerge at domain edges \cite{Hicks2010,Curran2014}. 

\begin{figure*}[tb]
\begin{center}
\includegraphics[width=1.99\columnwidth]{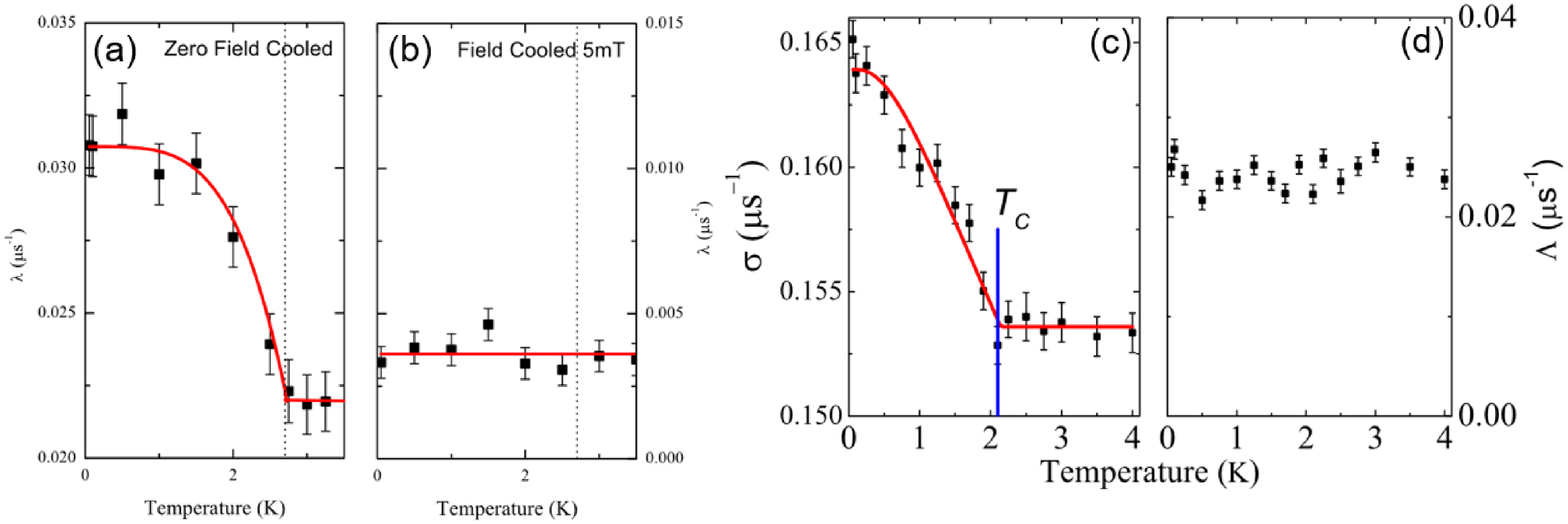}
\end{center}
\caption{Lorentzian relaxation rate $\lambdazf$ obtained from fitting $\mu$SR spectra of LaNiC$_2$ in (a) zero-field, and (b) a longitudinal field of 5~mT. In zero-field there is a clear increase of $\lambdazf$ correlated with $T_c$, which disappears in the longitudinal field. Reproduced with permission Ref. \onlinecite{Hillier2009}. Copyright 2009 by the American Physical Society. From the analysis of the zero-field $\mu$SR of LaNiGa$_2$, the temperature dependence of the (c) the Gaussian relaxation rate $\sigmazf$, and (d) the Lorentzian relaxation rate $\Lambda$ are displayed. In this case  $\sigmazf$ shows a clear increase below $T_c$, while $\Lambda$ remains unchanged. Reproduced with permission Ref. \onlinecite{Hillier2012}. Copyright 2012 by the American Physical Society.
 }
\label{LaNiC2Fig}
\end{figure*}

The long held understanding of chiral $p$-wave superconductivity in Sr$_2$RuO$_4$ has been challenged by the recent findings of a decrease in the Knight shift below $T_c$ in $^{17}$O NMR measurements for in-plane fields \cite{Pustgow2019}, which is contrary to both earlier NMR studies \cite{Ishida1998}, as well as the expected temperature-independent behavior  for chiral $p$-wave pairing. While triplet states with $\mathbf{d(k)}\perp c$ could not be completely excluded, these are not expected to exhibit TRS breaking. Moreover, this surprising result has been corroborated by subsequent spin-susceptibility measurements using polarized neutron scattering, where a drop of around $34\%$ was detected at low temperatures compared to the value above $T_c$ \cite{Petsch2020}. Spin-singlet superconductivity which preserves time-reversal symmetry was proposed  from mapping the momentum dependence of the gap structure using quasiparticle interference imaging, where it was suggested that the results are most consistent with $d_{x^2-y^2}$ symmetry, similar to that of the cuprates \cite{Sharma2019}. Such a lack of time-reversal symmetry breaking was concluded from a recent study of Josephson junctions formed between Sr$_2$RuO$_4$  and Nb, due to the invariance of the Josephson critical current upon reversing the current and field directions \cite{Kashiwaya2019}. On the other hand, as mentioned above the $\mu$SR evidence for TRS breaking has been backed up by optical Kerr effect measurements~\cite{Xia2006}. Moreover, a recent $\mu$SR study of Sr$_2$RuO$_4$  under uniaxial stress again confirmed the enhancement of the $\mu$SR relaxation rate below $T_c$ in unstressed samples \cite{Grinenko2020}. Interestingly, the application of uniaxial stress appears to decouple the onset of TRS breaking from the superconducting transition, where there is a slight decrease of the onset temperature with increasing stress, while $T_c$ increases. Evidence for a two-component order parameter has also been reported from two studies, based on a discontinuity in the shear elastic modulus $c_{66}$ at $T_c$ \cite{Ghosh2020,Benhabib2020}. It was pointed out that both a chiral $d_{xz}+id_{yz}$ state, and an accidentally degenerate $d_{x^2-y^2}+ig_{xy(x^-y^2)}$ state corresponding to a mixture of representations, are compatible with the requirements of a multicomponent singlet order parameter with broken TRS. However, further studies are still required to determine the nature of the pairing state, and resolve the apparent discrepancies between the results of different measurement techniques.

It should be noted  that TRS breaking is by no means an ubiquitous feature of strongly correlated unconventional superconductors (Table~\ref{tab:unconv}). In addition to the aforementioned stoichiometric UBe$_{13}$ \cite{Heffner1990}, a lack of evidence for TRS breaking is also found from $\mu$SR measurements of the heavy fermion superconductors Ce(Co,Ir)In$_5$ \cite{Higemoto2002} and CeCu$_2$Si$_2$ \cite{Feyerherm1997}.  Meanwhile, zero-field $\mu$SR measurements do not reveal evidence for broken TRS in the superconducting states of the high-$T_c$ cuprate superconductors YBa$_2$Cu$_3$O$_7$ (YBCO) and Bi$_2$Sr$_{2-x}$La$_x$CuO$_{6+\delta}$ \cite{Kiefl1990,Russo2007,Saadaoui2013}, as expected for $d_{x^2-y^2}$ symmetry (although evidence for a spontaneous magnetization below $T_c$ was reported from SQUID measurements of YBCO films \cite{Carmi2000}). In the case of the Fe-based superconductors, optimally doped Ba$_{1-x}$K$_x$Fe$_2$As$_2$ ($x\approx0.6$) has been widely suggested  to have a nodeless sign-changing $s_{\pm}$ pairing state \cite{Mazin2008,Ding2008,Nakayama2009,Hiraishi2009} while heavily hole doped KFe$_2$As$_2$ was proposed to be a nodal $d$-wave superconductor \cite{Fukazawa2009,Dong2010,Reid2012,Maiti2011}. It was predicted that the crossover between these pairing states with different symmetries should be via an intermediate $s+id$ state with broken TRS \cite{Lee2009}. However, apparently conflicting results are found experimentally, where a zero-field $\mu$SR study of a number of dopings in the range $x=0.5-0.9$ found no evidence for TRS breaking \cite{Mahyari2014}. Meanwhile, another $\mu$SR study of a crystal with doping $x\approx0.73$, which was ion-irradiated so as to create lattice symmetry breaking defects, showed evidence for TRS breaking in the superconducting state \cite{Grinenko2017}. The different results from the previous study was attributed to the $s+id$ state existing only over a narrow doping range. On the other hand, subsequent studies suggested that $s$-wave pairing symmetry is preserved across the entire Ba$_{1-x}$K$_x$Fe$_2$As$_2$ phase diagram, even in the nodal states of the heavily hole doped samples \cite{Okazaki2012,Cho2016}.

\subsection{Rare-earth based filled-skutterudites}

The filled-skutterudites $MT_4X_{12}$ ($M$ = rare earth or alkaline earth, $T$ = transition metal, and $X$ = Ge, P, As, Sb) are a large family of materials with fascinating properties, in which several compounds become superconducting at low temperature \cite{Shirotani1997,Takeda2000,Bauer2002,Bauer2007,Gumeniuk2008}. Among these, the heavy fermion superconductor PrOs$_4$Sb$_{12}$ is unique in that superconductivity is likely mediated by quadrupolar fluctuations \cite{Bauer2002}. On the other hand, other filled skutterudite superconductors behave more like conventional s-wave superconductors \cite{Tee2012,Chia2004,Pfau2016,Shimizu2007}, lacking evidence for strong electronic correlations. TRS breaking has been found from zero-field $\mu$SR measurements of PrOs$_4$Sb$_{12}$ \cite{Aoki2003} and PrPt$_4$Ge$_{12}$ \cite{Maisuradze2010}. In the case of PrOs$_4$Sb$_{12}$, this conclusion was also supported by Kerr effect measurements \cite{Levenson2018}. The TRS breaking appears to be related to the presence of Pr$^{3+}$ ions, since the size of the increase in $\sigma_{\rm ZF}$ below $T_c$ is reduced by La-doping, and TRS breaking is not detected in the purely La-based compounds \cite{Shu2011,Zhang2019}. Furthermore, a more rapid suppression of TRS breaking is observed when replacing Os with Ru atoms in PrOs$_4$Sb$_{12}$ \cite{Shu2011}. 

The superconducting properties of $MT_4X_{12}$  have been intensively studied, in particular for the compounds PrOs$_4$Sb$_{12}$ and Pr$_{x}$Pt$_4$Ge$_{12}$, but the nature of their order parameter remains to be established. A number of measurements, including  measurements of the lower critical field, thermal conductivity and $\mu$SR suggest fully gapped multiband superconductivity for these compounds \cite{Cichorek2005,Chandra2012,Hill2008,Seyfarth2005, Seyfarth2006,Shu2009}. However, there are also signatures of unconventional superconductivity in PrOs$_4$Sb$_{12}$, such as the absence of a coherence peak in $1/T_1$ below $T_c$ \cite{Kotegawa2003}, as well as evidence for point nodes from the penetration depth \cite{Chia2003} and angular dependent thermal transport \cite{Izawa2003}. Substitution of either Pr by La, or Os by Ru in PrOs$_4$Sb$_{12}$ leads to the suppression of the nodal behavior, suggesting a possible change of the pairing state in the doped compounds \cite{Shu2011,Zhang2019,Yogi2003}. Similar debate also exists for PrPt$_4$Ge$_{12}$. The early measurements of the specific heat and $\mu$SR suggest nodal superconductivity for PrPt$_4$Ge$_{12}$ \cite{Maisuradze2009}. Later measurements of the London penetration depth \cite{Zhang2013,Zhang2015b}, specific heat, thermal transport \cite{Pfau2016}, NQR \cite{Kanetake2010} and photoemission spectroscopy \cite{Nakamura2012} demonstrate that both PrPt$_4$Ge$_{12}$ and LaPt$_4$Ge$_{12}$ behave more like conventional BCS superconductors with a fully opened energy gap. The smooth evolution of the superconducting transition between the Pr-based end compounds with TRS breaking, and the time reversal preserving La-based materials, is difficult to account for and requires further studies.

\subsection{TRS breaking in fully gapped superconductors}

In recent years there have been a number of reported cases of materials with evidence for TRS breaking in the superconducting state,  which appear to be quite distinct from the aforementioned examples in strongly correlated SCs. The properties of these systems are also in \tab{tab:trsbsc}. Besides there being a lack of evidence for an underlying correlated state in many cases, the superconducting properties generally appear to behave similar to conventional $s$-wave SCs, i.e. the superconducting gap is fully open over the whole Fermi surface. This leads to the question of whether the origin of TRS breaking is from an unconventional superconducting state arising from a pairing mechanism other than the electron-phonon mechanism of BCS theory, or if such behavior can be realized via conventional pairing.

\subsubsection{LaNiC$_2$ and LaNiGa$_2$}

Evidence for TRS breaking has been found from zero-field $\mu$SR measurements of the SCs LaNiC$_2$ ($T_{\rm c} = 2.7$~K)~\cite{Hillier2009} and LaNiGa$_2$ ($T_{\rm c} = 2.1$~K)~\cite{Hillier2012}. These materials crystallize in different, but related orthorhombic crystal structures. LaNiGa$_2$ crystallizes in a centrosymmetric structure with space group $Cmmm$ (point group $D_{2h}$) \cite{Zeng2002}, whereas LaNiC$_2$ has a noncentrosymmetric space group $Amm2$ (point group $C_{2v}$), where inversion symmetry is broken within the Ni-C layer lying half way between the $A$-faced centers \cite{Lee1996}. The TRS breaking is manifested by an abrupt increase in the relaxation rate of the asymmetry in the zero-field $\mu$SR spectra measured below $T_{\rm c}$, while above $T_{\rm c}$, the spectra are temperature independent (Fig.~\ref{LaNiC2Fig}). Upon analysis using Eq.~\ref{AsymFit}, it was found that the  increase below $T_{\rm c}$ is in  $\lambdazf(T)$ for LaNiC$_2$, but $\sigmazf(T)$ for LaNiGa$_2$. Since however, the effect is destroyed in LaNiC$_2$ by the application of a small longitudinal field  of 5~mT, the behavior for both compounds was ascribed to the spontaneous onset of weak static fields in the superconducting state. 

In LaNiC$_2$, the occurrence of TRS breaking is supported by magnetization measurements   using a SQUID magnetometer \cite{Sumiyama2015}. Here single crystalline LaNiC$_2$ was measured together with a reference SC Ta, where the latter was used to detect and cancel the stray field. The zero-field magnetization of LaNiC$_2$ was then determined by measuring the magnetic flux change upon cooling under these zero-field conditions. From the temperature dependence of the magnetization change  for the highest quality crystal, after subtracting the value at 3.5~K It was found that while no  features occur along the $a$-axis, there is a distinct increase of $\Delta M(T)$ along the $c$-axis. Moreover, when the sample direction is reversed, the sign of the magnetization change is also flipped, suggesting that this magnetization is intrinsic to the sample. While in TRS breaking SCs, the presence of differently orientated domains may be expected to lead to zero net magnetization (as opposed to the local magnetic fields detected using $\mu$SR), the authors suggest that the lack of equivalence between the [$001$] and [$00\bar{1}$] directions in the noncentrosymmetric crystal structure may lead to pinning of the magnetization along this direction.

\begin{figure}[t]
\begin{center}
\includegraphics[width=0.7\columnwidth]{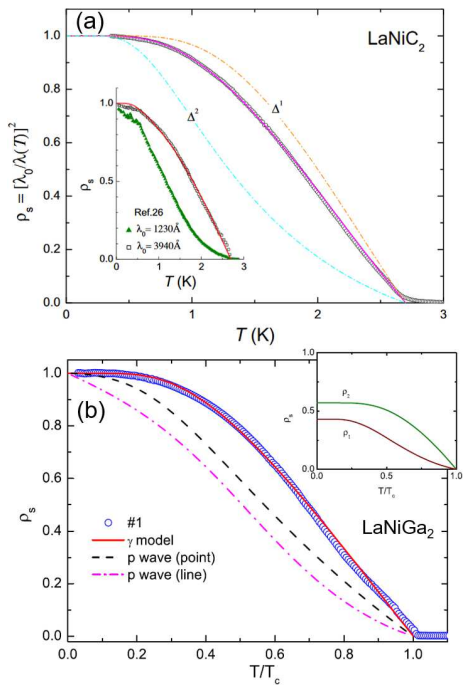}
\end{center}
\caption{ (a) Temperature dependence of the normalized superfluid density, derived from $\Delta\lambda(T)$ measured using the tunnel-diode oscillator based technique for (a) LaNiC$_2$, and (b) LaNiGa$_2$. The solid lines in both panels show the results from fitting using  nodeless two-gap models. Panel (a) is reproduced from Ref.~\onlinecite{Chen2013}, available under a Creative Commons Attribution 3.0 Unported (CC-BY) license; (b) is reproduced with permission Ref. \onlinecite{Weng2016}. Copyright 2016 by the American Physical Society.
 }
\label{LaNiC2PenFig}
\end{figure}

The superconducting gap structures of both materials have been probed via a number of methods. While initially a $T^3$ dependence of the specific heat was reported for LaNiC$_2$ \cite{Lee1996}, suggesting a line nodal gap, subsequent measurements at low temperatures reported the exponential behavior characteristic of a fully open gap \cite{Pecharsky1998,Chen2013}. This conclusion was corroborated by penetration depth measurements using the tunnel-diode oscillator based technique \cite{Chen2013}, as well as a nuclear quadrupole resonance (NQR) study of the spin-lattice relaxation rate \cite{Iwamoto1998}. Meanwhile fully-gapped superconductivity in LaNiGa$_2$ was reported from measurements of both the specific heat and magnetic penetration depth \cite{Zeng2002,Weng2016}. Moreover, from combining low temperature magnetic penetration depth measurements performed using the tunnel-diode oscillator based  method, specific heat and upper critical field results, it was found that the behaviors of both compounds are consistently well described by a  model with two nodeless gaps (Fig.~\ref{LaNiC2PenFig})  \cite{Weng2016,Chen2013}, indicating that nodeless two-gap superconductivity is another common feature of LaNiC$_2$ and LaNiGa$_2$. These experimental results were reconciled by the proposal of an internally-antisymmetric nonunitary triplet ground state, described in more detail in \sect{sec:novel-gs}.

\subsubsection{Noncentrosymmetric La$_7X_3$ and Re$X$}
\begin{figure*}[tb]
\begin{center}
\includegraphics[width=1.99\columnwidth]{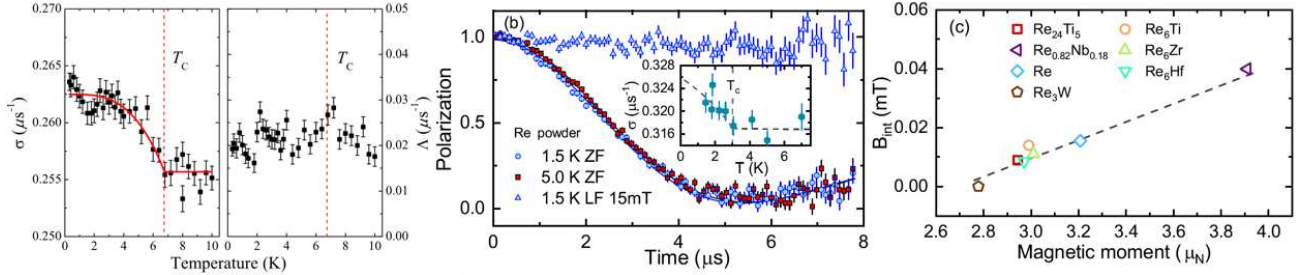}
\end{center}
\caption{(a) Temperature dependence of the Lorentzian and Gaussian relaxation rates obtained from zero-field $\mu$SR measurements of Re$_6$Zr, where the latter shows a clear increase below $T_c$ indicating TRS breaking. Reproduced from Ref.~\onlinecite{Singh2014}, available under a Creative Commons Attribution 3.0 License.  (b) $\mu$SR time spectra of elemental Re measured in zero-field above and below $T_c$, and in a longitudinal field of 15~mT in the superconducting state. The more rapid depolarization below $T_c$ indicates the presence of TRS breaking of elemental Re, and as shown in the inset this onsets below $T_c$. (c) Plot of the internal field estimated from the increase of $\sigmazf$ below $T_c$ as a function of the nuclear moment calculated from the nuclear moments of the constituent elements and their relative fractions. A clear positive correlation is observed between the two quantities. Both (b) and (c) are reproduced with permission from Ref.~\onlinecite{ShangPRL2018}. Copyright 2018 by the American Physical Society.
 }
\label{ReXFig}
\end{figure*}

In noncentrosymmetric SCs, antisymmetric spin-orbit coupling (ASOC) can give rise to a mixed singlet-triplet pairing state. As discussed in \sect{sec:novel-gs}, however, the low symmetry of the orthorhombic point group of LaNiC$_2$ means that even if inversion symmetry is broken, the ASOC cannot give rise to mixed singlet-triplet pairing states which break time-reversal symmetry at $T_c$. However, TRS breaking has been found in several other noncentrosymmetric SCs with higher symmetries, which opened up the possibility for such singlet-triplet mixing.

\begin{figure}[t]
\begin{center}
\includegraphics[width=0.8\columnwidth]{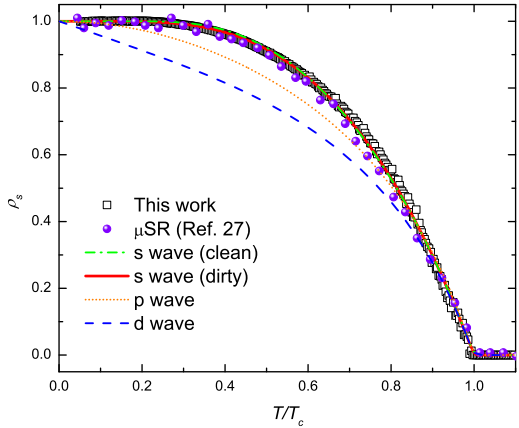}
\end{center}
\caption{Temperature dependence of the superfluid density of Re$_6$Zr, derived from measurements of the magnetic penetration depth using the tunnel-diode oscillator based method. The superfluid density determined from $\mu$SR measurements are also displayed \cite{Singh2014}, where close agreement is found from the two methods. The data are well described by single gap isotropic $s$-wave models for both the clean and dirty limits. Reproduced  with permission from Ref.~\onlinecite{Pang2018}. Copyright 2018 by the American Physical Society.
 }
\label{ReZrSuperfl}
\end{figure}

 La$_7$Ir$_3$ and La$_7$Rh$_3$ crystallize in the noncentrosymmetric hexagonal Th$_7$P$_3$ type-structure with $T_c=2.3$ and 2.65~K, respectively. Zero-field $\mu$SR on  La$_7$Ir$_3$ revealed a clear increase of  $\lambdazf(T)$ below $T_c$, similar to the case of LaNiC$_2$, while $\sigmazf$ shows little change \cite{Barker2015}. Meanwhile from $\mu$SR measurements in an applied transverse field, the temperature dependence of the superfluid density derived from the magnetic penetration depth $\lambda(T)$ is very well described by a single isotropic gap,  with a magnitude slightly larger than that of weak coupling BCS theory \cite{Barker2015}. The good accordance with single gap $s$-wave superconductivity is also supported by the analysis of the specific heat \cite{Li2018}. Furthermore, the estimated $T_c$ based on a  first principles calculation of the phonon frequencies and electron phonon-coupling constant is very close to the experimentally observed value, giving a strong indication of a conventional electron-phonon pairing mechanism \cite{Li2018}. Evidence for TRS breaking was also found in  La$_7$Rh$_3$, where the penetration depth could  be described by a single band $s$-wave model \cite{Singh2018a}.

Another large class of materials where evidence for TRS breaking superconductivity is found is in Re-based alloys Re$X$ with the  $\alpha$-Mn type cubic structure (space group $I\bar{4}3m$), which were first reported by Matthias \textit{et al.} nearly 60 years ago \cite{Matthias1961,Blaugher1961}. The crystal structure has a large unit cell with 58 atoms and four distinct crystallographic sites. These systems are generally non-stoichiometric with intrinsic site disorder, especially on the two sites with 24$g$ Wyckoff positions. TRS breaking below $T_c$ was initially found from $\mu$SR measurements of Re$_6$Zr \cite{Singh2014}, as displayed in Fig.~\ref{ReXFig}. Here the symmetry analysis for the cubic point group shows that TRS breaking singlet-triplet mixing pairing states below $T_c$ are  permitted. TRS breaking was subsequently reported in several other Re$X$ systems including Re$_6$Hf \cite{Singh2017}, Re$_6$Ti/Re$_{24}$Ti$_5$ \cite{Singh2018,Shang2018} and Re$_{0.82}$Nb$_{0.18}$ \cite{ShangPRL2018}. In all cases, the TRS breaking is manifested by the increase of $\sigmazf$ below $T_c$. Moreover, while some studies have found evidence for two-gap superconductivity in Re$X$ \cite{Cirillo2015,Parab2019}, most measurements were generally well accounted for by a single gap $s$-wave model \cite{Singh2014,Shang2018,Singh2017,Singh2018,ShangPRL2018,Lue2013,Chen2013b,Chen2016,Khan2016,Matano2016,Mayoh2017,Pang2018,Huang2018}. An example of this is displayed in Fig.~\ref{ReZrSuperfl}, where it can be seen that the superfluid density derived from penetration depth measurements of single crystals of Re$_6$Zr using the tunnel diode oscillator based method are well described using a model with a single isotropic gap \cite{Pang2018}. Furthermore, these data are highly consistent with the superfluid density determined from TF-$\mu$SR results \cite{Singh2014}.

A very surprising result was however reported in Ref.~\onlinecite{ShangPRL2018}, which shows that the element rhenium also exhibits the signatures of TRS breaking in superconducting state [Fig.~\ref{ReXFig}(b)]. Rhenium has a simple centrosymmetric crystal structure, and this therefore suggests that the breaking of inversion symmetry and the accompanying ASOC are not crucial ingredients for realizing TRS breaking in this series. This rather suggests that the local electronic structure of Re is important. This is consistent with $\alpha$-Mn type SCs which do not contain Re but still have sizeable ASOC, such as Nb$_{0.5}$Os$_{0.5}$ and  Mg$_{10}$Ir$_{19}$B$_{16}$, being found   not to exhibit TRS breaking \cite{Aczel2010,Singh2018b}. Furthermore, as shown in Fig.~\ref{ReXFig}(c)  a clear correlation is found between the magnitude of the internal field emerging below $T_c$, which can be estimated from the increase of $\sigmazf$, and the size of the nuclear moments $\mu_n$ corresponding to $\mu_{\rm n}=\sqrt{f_{\rm Re}\mu_{\rm n,Re}^2+f_{X}\mu_{\rm n,X}^2}$, where $\mu_{\rm n,Re}$ and $\mu_{\rm n,X}$ are the nuclear moments of Re and $X$ elements, and $f$ are the relative fractions. Since  $\mu_{\rm n,Re}$ is relatively large, the systems with larger Re content generally show a larger internal field, while in less Re rich Re$_3$W, TRS breaking cannot be detected \cite{Biswas2012}. However the origin of this correlation  is difficult to account for, and this requires further experimental and theoretical study.

\subsubsection{Weakly-correlated centrosymmetric superconductors}

Although most of the recent examples of superconductors with broken TRS have been  noncentrosymmetric systems, the role played, if any, by the broken inversion symmetry in giving rise to this phenomenon remains to be established. The crystal structure of SrPtAs is globally centrosymmetric, with the hexagonal space group $P6_3/mmc$, and it becomes superconducting below $T_c=2.4$~K \cite{Nishikubo2011}. However, the structure consists of two Pt-As layers stacked alternately along the $c$-axis, where inversion symmetry is broken locally within the layers. Therefore if the coupling between Pt-As layers is weak, novel superconducting pairing states may be realized \cite{Fischer2011}, similar to the case where inversion symmetry is globally absent. TRS was found to be broken below $T_c$ in SrPtAs from an increase of $\lambdazf$ below $T_c$ \cite{Biswas2012}. Despite a number of novel theoretical proposals such as chiral $d$-wave and $f$-wave pairings \cite{Goryo2012,Wang2014,Fischer2014}, Knight shift measurements indicate the presence of singlet pairing \cite{Matano2014}, and probes of the gap structure can be accounted for with  a nodeless isotropic gap~\cite{Biswas2012,Matano2014,Bruckner2014,Landaeta2016}. Note that such alkaline-earth based equiatomic ternary compounds
crystallize in several different structures. CaPtAs is a noncentrosymmetric superconductor with a tetragonal structure and a $T_c$ of 1.47~K \cite{Xie2020}, where evidence for broken TRS is also found from $\mu$SR measurements and there is likely a nodal gap structure \cite{ShangUnPub}.

The  $R_5$Rh$_6$Sn$_{18}$ ($R$=Lu, Y, Sc) SCs have a centrosymmetric caged structure \cite{Remeika1980}. Here the tetragonal structure with space group $I4_1/acd$ corresponds to a distortion of a cubic structure, where the periodicity of the lattice doubles along the $c$-axis. The signatures of TRS breaking were found for all three compounds from zero-field $\mu$SR measurements \cite{Bhattacharyya2015,Bhattacharyya2015A,Bhattacharyya2018}, as manifested in an increase of $\lambdazf$ below $T_c$. On the basis of a symmetry analysis, two potential unconventional TRS breaking states were proposed, a $d+id$ singlet state and a non-unitary triplet state \cite{Bhattacharyya2015}. The former has two point nodes and a line node on a three-dimensional Fermi surface, while the latter has only point nodes. On the other hand, the superfluid density extracted from transverse field $\mu$SR measurements for $R$=Lu and Y saturates at low temperature, and is well described by a single isotropic gap \cite{Bhattacharyya2015,Bhattacharyya2015A}. Moreover, nodeless superconductivity was deduced for these two compounds from the absence of a residual contribution to the thermal conductivity at low temperatures \cite{Zhang2015a}, while  the specific heat of Sc$_5$Rh$_6$Sn$_{18}$ is also well described by an $s$-wave model \cite{Kase2011b}. However, in the case of Y$_5$Rh$_6$Sn$_{18}$ evidence for gap anisotropy was found from a four-fold oscillation of the specific heat coefficient in the $ab$-plane \cite{Kase2014}, and a non-linear field dependence of the same quantity \cite{Kase2011}, suggesting a possible deviation from isotropic $s$-wave superconductivity.

\section{Theoretical analysis}
\mylabel{sec:GL-theory}

In this section, we outline theoretical efforts to understand the order parameter symmetry and underlying pairing mechanisms of the TRS breaking SCs discussed in the previous section. The starting point is the group-theoretical formulation of Ginzburg-Landau (GL) theory~\cite{Volovik1996,Annett1990,Sigrist1991}. This very general technique allows us to obtain important information about the superconducting ground state using only general symmetry properties of the material, that is, without requiring information about the underlying pairing mechanism. Once this has been achieved, the experimentally observable properties of the symmetry-allowed ground states can be investigated by a generalized BCS-type mean-field approach within the Bogoliubov-de Gennes formalism~\cite{Sigrist1991}. This second step may require  a model for the pairing interaction. We describe applications of such approaches to the materials discussed above, emphasizing how experimental data, band structure and such theories can in principle lead to a complete understanding of the properties of the TRS-breaking superconducting ground states. We also discuss the difficulties that are encountered in practice in trying to carry out such a programme.

\subsection{Group theoretical formulation of the Ginzburg-Landau theory}
Here we review the group theoretical formulation of the GL theory used to constrain possible forms of superconducting order parameters using the underlying symmetries of the crystal structure without requiring the knowledge of the pairing mechanisms. We use specific examples of TRS breaking superconducting materials to illustrate what can be learned about possible order parameters based on these general considerations. In particular, we discuss the conditions under which TRS breaking implies symmetry-required nodes on the Fermi surface in view of their respective band structures and the constraints on the pairing state imposed by spin-orbit coupling depending on the presence of inversion symmetry in the SCs. 

Landau theory~\cite{Toledano1987} is a phenomenological theory of continuous phase transitions. It has been immensely successful in accurately predicting the qualitative features of a wide range of phase transitions featuring change in symmetry. The basic assumption of the theory is that there is a continuous transition at a given transition temperature ($T_c$), between symmetrically distinct low temperature and high temperature phases. The aim is to establish a relation between the symmetries of the two phases and find relevant physical thermodynamic quantities which change anomalously across the transition. This is achieved by introducing the concept of an order parameter and a thermodynamic potential. The change in symmetry across the phase transition is quantified by the order parameter, a thermodynamic quantity (e.g. magnetization of a ferromagnet) which is zero in the symmetric (disordered) high temperature phase and nonzero in the ordered phase at low temperatures. Since the transition is assumed to be continuous, the absolute size of the order parameter can be assumed to be small near $T_c$ such that the thermodynamic potential can be expressed as a Taylor expansion of the order parameter components and only one irreducible degree of freedom participates in determining the symmetry breaking at $T_c$. Important consequences are discontinuous changes in physical quantities related to second-order derivatives of the thermodynamic potential at $T_c$ and the absence of co-existent regions of two phases.    

We now focus on the case of a superconducting phase transition. From the point of view of GL theory, the superconducting phase transition is a continuous phase transition accompanied by spontaneous gauge symmetry breaking at $T_c$. The latter leads to the rigidity of the phase of the electron fluid which in turn gives rise to the classic properties of SCs: zero resistance, Meissner effect and the Josephson effect~\cite{Anderson2018}.

\subsection{Normal state symmetry group}
In discussing the symmetry properties of the possible superconducting order parameters we first need to consider the generic symmetry properties of the material. In the normal state, the system must be invariant under the normal state symmetry group~\cite{Volovik1996,Annett1990,Sigrist1991} defined as
\beq
\mathcal{G} = G \otimes U(1) \otimes \mathfrak{T}.
\eeq
Thus $\mathcal{G}$ is a direct product (represented by $\otimes$) group of the group $G$ of crystalline symmetries with rotations in spin space, the gauge symmetry group $U(1)$ and the group of time reversal operations $\mathfrak{T}$. When spin-orbit coupling (SOC) is not important and the normal state is nonmagnetic, we can write $G = G_c \otimes SO(3)$ where G$_c$ is the space group of the crystal and $SO(3)$ is the group of rotations in spin space in 3D.

Landau theory states that the possible symmetries of the order parameter just below $T_c$ are determined by the irreducible representations (irreps) of $G$. Near $T_c$, the GL free energy can be approximated by a finite Taylor expansion in the order parameter components as the magnitude of the order parameter itself is small. The free energy thus expressed must be invariant under $G$. Remarkably, this is enough to constrain the possible symmetries of the order parameter. SCs which only break the $U(1)$ symmetry are called conventional SCs and those with additional broken symmetries other than $U(1)$ are termed, in this context, unconventional SCs.

We will consider superconducting instabilities that are uniform throughout the system. In this case, the possible superconducting order parameters can be constructed by considering only the symmetries of a single unit cell. For a crystal with symmorphic space group, the symmetry of a unit cell is described by the point group (containing rotations and reflections) of the underlying Bravais lattice. In this case, we can write $G_c = G_0 \otimes T$ where $T$ is the translation group of the crystal and $G_0$ is the crystalline point group. However, for a crystal with nonsymmorphic space group symmetries the symmetry operations within a unit cell include nonsymmorphic operations such as screw axes (rotation about an axis with a fractional translation, i.e. by a fraction of the primitive lattice vector, parallel to the axis of rotation) and/or glide planes (reflection in a plane followed by a fractional translation parallel with that plane). These are in addition to some of the regular point symmetry operations. 

The presence of nonsymmorphic symmetries has important consequences for the gap structure of a material which include the violation of  Blount's theorem~\cite{Blount1985}. Blount's theorem states that no symmetry protected line nodes are allowed in an odd-parity SC. Although symmorphic crystals obey this theorem, recent studies~\cite{Yanase2016,Sumita2017,Micklitz2017,Sumita2018} have shown that nonsymmorphic symmetries can lead to symmetry protected line nodes even in odd-parity SCs. The essence is that the nonsymmorphic symmetries can cause additional symmetry-required nodes on the Brillouin zone faces along high-symmetry directions. These nodes will affect the thermodynamic properties of the system when Fermi surfaces touch the Brillouin zone edges \emph{and} are not ``open'' along those high-symmetry directions. Barring these symmetry-required nodes, the overall symmetry of the order parameter can still be determined by considering the underlying point group operations of the crystal. This is the approach we take in the following discussions. 

\subsection{Construction of the GL free energy}
Under the general assumptions of  Landau theory, i. e. the existence of an order parameter and its continuous change as a function of temperature across $T_c$, we can construct a generic form valid near $T_c$ of the GL free energy as a Taylor expansion of the order parameter. In this regime, the order parameter and its spatial variation is assumed to be small.  Taking the order parameter to consist of a set of complex numbers $\{\Delta_i\}$ which will vary continuously as functions of $T$ and vanish when $T>T_c$, the generic form of the GL free energy~\cite{Annett1990,Sigrist1991} in the absence of magnetic fields is given by
\bea
f = f_0 &+& \sum_{i,j} \Delta^*_i \alpha_{i,j} \Delta_j + \sum_{i,j,k,l} \Delta^*_i \Delta^*_j \beta_{i,j,k,l} \Delta_k \Delta_l \non\\
&+& \sum_{i,j,k,l} \partial_i\Delta^*_j {\mathcal{K}}_{i,j,k,l} \partial_k \Delta_l + \ldots
\eea
Here, $f_0$ is the free energy of the normal state, $\alpha_{i,j}$ are the elements of the inverse pairing susceptibility matrix $\hat{\alpha}$, $\beta_{i,j,k,l}$ are the elements of the $\hat{\beta}$ tensor characterizing the fourth order term and ${\mathcal{K}}_{i,j,k,l}$ are the elements of the $\hat{\mathcal{K}}$ tensor characterizing the spatial variation of the free energy. The free energy has the same symmetry as an effective action or an effective Hamiltonian describing the normal state and must be invariant under the transformations of the normal state symmetry group $\mathcal{G}$. As a result, the way the order parameter changes under the operations in $\mathcal{G}$ implies that the elements of $\hat{\alpha}$, $\hat{\beta}$ and $\hat{\mathcal{K}}$ are constrained by symmetry. For example, 1) the free energy must be real and invariant under TRS which constrains $\hat{\alpha}$ to be a real symmetric matrix and $\hat{\beta}$ to be a real tensor, 2) $\beta_{i,j,k,l} = \beta_{j,i,k,l} = \beta_{i,j,l,k} = \beta_{j,i,l,k}$, 3) free energy has to be invariant under $G$ which implies $\hat{\alpha} = \hat{R}^\dagger_g \hat{\alpha} \hat{R}_g$ $\forall g \in G$ with $\hat{R}_g$ being a matrix representation of the element $g$.

Representation theory of groups can now be used to block-diagonalize the $\hat{\alpha}$ matrix with blocks of the same dimensions as those of the irreps of $G$. Each block can further be transformed, using a suitable basis for the irrep of dimension $d$, into a number times the identity matrix of order $d$. Thus a $d$-dimensional irrep produces a $d$-fold degenerate eigenvalue. These eigenvalues of the $\hat{\alpha}$ matrix are of special significance since they correspond to all the possible symmetry allowed channels of superconducting instabilities in the system. 

At high temperature the superconducting state is unstable, leading to all the eigenvalues being positive definite. But as the temperature is lowered, they can change sign. Assuming the eigenvalue $\alpha_1$ changes sign first, we can take the form $\alpha_1 = a_0 (T-T_c)$ with $a_0$ being a positive real number. The analyticity of $\alpha_1$ stems from the basic underlying assumption of continuous change of the order parameter as a function of temperature. The instability corresponding to $\alpha_1$ will have the highest $T_c$ and will characterize the superconducting properties of the ground state of the system.

We consider the instability with the highest $T_c$ to correspond to the $d$-dimensional irrep $\Gamma$ of $G$. Then the corresponding order parameter of the system close to $T_c$ can be written as
\beq\mylabel{eqn:gap_function}
\hat{\Delta}(\bk) = \sum_{m=1}^{d} \eta^{\Gamma}_m \Delta^{\Gamma}_m (\bk)
\eeq 
where $\eta^{\Gamma}_m$ are the complex amplitudes of the order parameter $\hat{\Delta}$, also called the gap matrix or the pairing potential, corresponding to the basis functions $\Delta^{\Gamma}_m (\bk)$ of $\Gamma$. The thermodynamic properties of the state are then completely described by the set $\{\eta^{\Gamma}_m\}$ and thus those amplitudes can be used as an alternative description of the order parameter. In the basis function space $\{\Delta^{\Gamma}_m (\bk)\}$, these numbers transform as: $\mathcal{T}\eta = \eta^*$ and $\mathcal{C}\eta = e^{i\phi} \eta$ where $\mathcal{T}$ is the TRS operator and $\mathcal{C}$ is the gauge transformation operator with $\phi$ being a phase.

We now restrict our discussion to the particular irrep $\Gamma$ and write the free energy as a Taylor expansion in $\{\eta_m\}$ (dropping the $\Gamma$ label):
\bea
f = f_0 &+& a_0 (T-T_c) \sum_m |\eta_m|^2 + \sum_{ijkl} \beta_{ijkl} \eta^*_i \eta^*_j \eta_k \eta_l \non\\
&+& \sum_{ijkl} \mathcal{K}_{ijkl} \partial_i \eta^*_j \partial_k \eta_l + \ldots
\eea
For overall stability, the fourth order term in the free energy must be positive definite. The terms corresponding to a particular order in the free energy can now be constructed by constructing invariant polynomials~\cite{Volovik1996,Annett1990,Sigrist1991} of that order corresponding to the particular irrep of the symmetry group. Thus the symmetry constraints result in a fourth order term which depends on a few, material-dependent parameters $\{\beta_n\}$. By minimizing the free energy for $T<T_c$ with respect to all the complex variables $\{\eta_i\}$, all the distinct superconducting states that can exist just below $T_c$ for an instability in the channel $\Gamma$ can now be obtained and a phase diagram showing which of these states is realized depending on the values of the GL parameters can be constructed. In each of the phases there will be a particular relation between the different $\eta_i$ coefficients. If their complex phases differ by more than a mere change of sign, then we have a state with broken TRS. Evidently this requires a degenerate instability channel, i.e. one with $d>1$.

It is worth noting that following the above procedure we may still find that some of the states in a given channel are degenerate. The degeneracy can be lifted~\cite{Sigrist1991} by the effects of crystal field splitting, SOC and/or strong coupling. Mostly, the fourth order term in the free energy is sufficient to get non-degenerate superconducting states when crystal field, SOC or strong coupling effects are taken into account. However, in some cases, a few spurious degeneracies can still remain requiring the need to consider higher order terms~\cite{Sigrist1991} in the free energy. We will only consider the free energy up to fourth order in the subsequent discussions. One important point to note is that classifying the superconducting states in this manner gives emphasis only on the symmetry of the order parameter and not on its specific form.

\subsection{Structure of the order parameter}
In determining the structures of the possible order parameters, it is important to consider the effect of SOC on the normal state band structure of the material. If SOC leads to significant band splitting near the Fermi level, most probably it will be important in determining the properties of the superconducting ground state. If we consider BCS-type pairing in an effective single band picture, the Cooper pairs can only be of two types: spin singlet and spin triplet. These two kinds of pairing, singlet and triplet, can be distinguished by their different behaviors under rotations in spin space in the absence of SOC or when the effect of SOC can be neglected. But when SOC is finite and can not neglected, spin rotation and space rotation can not be separated. Then, since the overall Cooper pair wave function has to be antisymmetric, only parity ($\hat{P}$) of the Cooper pair wave function can distinguish between spin singlet (even parity) and spin triplet (odd parity) states.

The above argument applies to any centrosymmetric SC. However, a noncentrosymmetric material  lacks inversion symmetry. As a result, parity is not a well defined symmetry in this case. The crucial difference between centrosymmetric and noncentrosymmetric SCs is that for noncentrosymmetric SCs the irreps of $G_0$ do not have distinct symmetries under inversion and thus each of them are compatible with both singlet and triplet pairing. This results in the admixture of singlet and triplet pairing within the same superconducting state of a noncentrosymmetric material  with strong SOC~\cite{Smidman2017}. In contrast, centrosymmetric SCs  have either purely singlet or purely triplet pairing even with strong SOC, distinguishable by their respective parities. Thus noncentrosymmetric SCs are in this sense special since SOC has more dramatic effect on them than in their centrosymmetric counterparts.

In the case of strong SOC, the single-particle states are no longer the eigenstates of spin and we need to label them rather by pseudospins. The pseudospin states are linear combinations of the spin eigenstates. However, the pseudospin states can be thought to be generated from the spin eigenstates by adiabatically turning on the SOC. Hence, the original spin eigenstates and the pseudospin states have one-to-one correspondence. If $\alpha$ and $\beta$ labels the pseudospin states then we can identify $\uparrow \equiv \alpha$ and $\downarrow \equiv \beta$. In such a system zero momentum Cooper pairs are formed from particles from energetically degenerate states. The pseudospin states $\ket{\bk,\alpha}$ and its time reversed partner $\mathcal{T}\ket{\bk,\alpha}=\ket{-\bk,\beta}$ (where $\mathcal{T}$ is the time-reversal operator) are paired for the case of even parity singlet pairing while four degenerate states $\ket{\bk,\alpha}$, $\mathcal{T}\ket{\bk,\alpha}=\ket{-\bk,\beta}$, $\mathcal{P}\ket{\bk,\alpha}=\ket{-\bk,\alpha}$ and $\mathcal{T}\mathcal{P}\ket{\bk,\alpha}=\ket{\bk,\beta}$ ($\mathcal{P}$ is the parity or the inversion operator) participate in pairing in the case of odd parity pairing. Since the pseudospin and the spin are closely related, the even parity states correspond to pseudospin singlet and the odd parity states correspond to pseudospin triplet states. These pseudospin superconducting states lead to important differences from their original spin counterparts which are apparent, for example, when considering a junction between two such materials with different SOC strengths~\cite{Sigrist1991}.

From the phenomenological GL theory of superconductivity~\cite{Tinkham2004,Coleman2015}, we know that the superconducting order parameter has the same symmetry as a Cooper pair wave function. Thus, the antisymmetry of fermionic wave function requires
\beq\mylabel{gap_antisymmtery}
\hat{\Delta}(\bk) = - \hat{\Delta}^T(-\bk).
\eeq 
Below $T_c$, it must have the full symmetry of the normal state and can be written in terms of the irreps of $G$ as shown in \eqn{eqn:gap_function}. For uniform superconducting instabilities and without any SOC, the structures of the order parameters are determined by the irreps of $G' = G_0 \otimes SO(3)$. If $\hat{\Gamma}^c$ and $\hat{\Gamma}^s$ are the irreps of $G_0$ and $SO(3)$ respectively, the irreps of $G'$ have the form
\beq
\hat{\Gamma} = \hat{\Gamma}^c \otimes \hat{\Gamma}^s.
\eeq 
Then if $\{\Gamma^c_m(\bk): m = 1, \ldots, n_c \}$ forms a basis for $\hat{\Gamma}^c$ with $n_c$ being its dimensionality and $\{\Gamma^s_n: n = 1, \ldots, n_s \}$ forms a basis for $\hat{\Gamma}^s$ with $n_s$ being its dimensionality, we can construct a basis for $\hat{\Gamma}$ as $\Gamma_{m,n}(\bk) = \Gamma^c_m(\bk)\Gamma^s_n$. The dimensionality of $\hat{\Gamma}$ is then given by $d = n_c n_s$. In the superconducting state, we can write the gap function from \eqn{eqn:gap_function} as
\beq
\hat{\Delta}(\bk) = \sum_{m=1}^{n_c} \sum_{n=1}^{n_s} \eta_{m,n} \Gamma^c_m(\bk)\Gamma^s_n.
\eeq

Thus, the order parameters in the singlet representations have the form
\beq
\hat{\Delta}_{\rm singlet}(\bk) = \sum_{m=1}^{n_c} \eta^s_{m} \Gamma^c_m(\bk)\Gamma^s_{\rm singlet} = \Gamma^s_{\rm singlet} \Delta_s(\bk) 
\eeq
requiring (by \eqn{gap_antisymmtery}) $\Gamma^s_{\rm singlet} = - \left(\Gamma^s_{\rm singlet}\right)^T$ and $\Gamma^c_m(-\bk) = \Gamma^c_m(\bk)$ which in turn implies $\Delta_s(-\bk) = \Delta_s(\bk)$, i.e. an even scalar function. The singlet case is thus called an even parity superconducting state. Similarly, in the triplet case, the order parameter can be written as
\beq
\hat{\Delta}_{\rm triplet}(\bk) = \sum_{m=1}^{n_c} \sum_{n=-1,0,1} \eta^t_{m,n} \Gamma^c_m(\bk)\Gamma^s_{triplet,n}
\eeq
requiring (by \eqn{gap_antisymmtery}) $\Gamma^s_{triplet,n} = \left(\Gamma^s_{triplet,n}\right)^T$ and $\Gamma^c_m(-\bk) = -\Gamma^c_m(\bk)$. The triplet case is thus called an odd parity superconducting state.

The spin dependence of the gap function can be compactly written using a general $2\times2$ matrix formalism as
\beq
\hat{\Delta}(\bk) = \begin{bmatrix}
\Delta_{\uparrow\uparrow} & \Delta_{\uparrow\downarrow}\\
\Delta_{\downarrow\uparrow} & \Delta_{\downarrow\downarrow}\\
\end{bmatrix}.
\eeq 
Then, $\hat{\Delta}(\bk)$ is called the gap matrix. Then the gap matrix for the singlet case is 
\beq
\hat{\Delta}_{\rm singlet}(\bk) = \Delta_s(\bk) i \sigma_y = \begin{pmatrix}0 & \Delta_s(\bk) \\ -\Delta_s(\bk) &0\end{pmatrix}\mylabel{eqn:op_singlet}
\eeq
where $\Delta_s(\bk) = \Delta_s(-\bk)$ is an even function of $\bk$. The gap matrix for the triplet case can be written as
\bea
\hat{\Delta}_{\rm triplet}(\bk) &=& \begin{bmatrix}
\Delta_{\uparrow\uparrow} & \Delta_0\\
\Delta_0 & \Delta_{\downarrow\downarrow}
\end{bmatrix} = \begin{bmatrix} 
-d_x + i d_y & d_z\\
d_z & d_x + i d_y
\end{bmatrix} \non\\
&=& (\bd(\bk).\hat{\pmb{\sigma}})i\sigma_y.\mylabel{eqn:op_triplet}
\eea
Here, $\hat{\pmb{\sigma}}$ is a vector with the three Pauli spin matrices as its components and $\bd(\bk) = -\bd(-\bk)$ is a complex odd vector function which transforms as a $3$D vector under rotation in the spin space. 

The $\bd(\bk)$-vector compactly describes the spin and angular momentum of the Cooper pairs and the nodal structure of the energy gap. In this case, it is instructive to define the vector
\beq
\bq(\bk) = i(\bd(\bk) \times \bd^*(\bk)).
\eeq
Note that $\bq(\bk)$ is a real even vector function of $\bk$. Then depending on the value of $\bq(\bk)$, there are two types of possible triplet pairing: $\bq(\bk) = 0$ is called a unitary triplet pairing state and $\bq(\bk) \neq 0$ is called a nonunitary triplet pairing state. The finite $\bq(\bk)$ corresponding to the non-unitary triplet pairing case is of special interest since it implies structural difference in pairing of up spins and down spins along different $\bk$-directions which in turn results in TRS breaking. In general, we have
\beq
\hat{\Delta}_{\rm triplet}(\bk) \hat{\Delta}^\dagger_{\rm triplet}(\bk) = |\bd(\bk)|^2\mathbb{1}_2 + \bq(\bk).\hat{\pmb{\sigma}}.
\eeq
This product plays an important role in distinguishing between the unitary and nonunitary types of pairing states.

For centrosymmetric SCs with strong SOC, the pseudospin singlet and pseudospin triplet order parameters have the same form as in \eqn{eqn:op_singlet} and \eqn{eqn:op_triplet} respectively. Whereas for noncentrosymmetric SCs with strong SOC, the superconducting state with mixed singlet and triplet components has the gap matrix
\beq
\hat{\Delta}(\bk) = [\Delta_s(\bk) + \bd(\bk).\hat{\pmb{\sigma}}]i\sigma_y.
\eeq 

We note that, in general, the singlet gap function $\Delta_s(\bk)$ can be written as
\beq
\Delta_s(\bk) = \sum_m c_m \mathcal{Y}_{l,m}(\hat{\bk}) \,\,\,\text{with \textit{l} = even}
\eeq
and the triplet $\bd$-vector can be written as
\beq
\bd(\bk) = \sum_{m,n=x,y,z} b_{m,n} \mathcal{Y}_{l,m}(\hat{\bk}) \hat{n}\,\,\,\text{with \textit{l} = odd}.
\eeq
Here, $\mathcal{Y}_{l,m}(\hat{\bk})$ are the spherical harmonics in the $\bk$-space, $c_m$ and $b_{m,n}$ are complex coefficients. Thus, these gap functions are given a nomenclature according to their respective angular momentum quantum number. In general, if the gap matrix satisfies $\hat{\Delta}(\delta \bk) = \delta^l \hat{\Delta}(\bk)$ for a real number $\delta$, then $l = 0$, $1$, $2$, $\ldots$ are called s-wave, p-wave, d-wave, $\ldots$ superconducting states respectively.

\section{Time reversal symmetry breaking order parameters}
\mylabel{sec:GL-theory-TRSB}
We now specialize to the case of SCs which spontaneously break TRS  in this section. We first note that TRS breaking in the superconducting ground state requires a degenerate instability channel corresponding to a multi-dimensional irrep of $\mathcal{G}$. This argument hinges on the fact that  a TRS breaking state with order parameter $\hat{\Delta}(\bk)$ under the TRS operation should transform to a new state with order parameter $\hat{\Delta}'(\bk) = \mathcal{T}\hat{\Delta}(\bk)$ where $\hat{\Delta}'(\bk)$ is not simply related by an overall phase to $\hat{\Delta}(\bk)$. The simplest possible way it can happen is by a multi-component order parameter with nontrivial phase difference between its components. The origin of the multi-component order parameter from a multidimensional irrep of $\mathcal{G}$ can however be very different, for example, it can correspond to the underlying point group $G_0$ (an example of this is the chiral $p$-wave state proposed for Sr$_2$RuO$_4$~\cite{Mackenzie2003}), the group of spin rotations $SO(3)$ (for example a nonunitary triplet state with equal spin pairing proposed for the cases of LaNiC$_2$ and LaNiGa$_2$~\cite{Hillier2009,Hillier2012,Weng2016,Quintanilla2010}) or the TRS group itself (for example the special case of the chiral point group $\mathcal{C}_4$ allowing for a loop-super current ground state~\cite{Ghosh2018}). 

We note that the classification scheme of the superconducting order parameters discussed in the previous section can lead to many symmetry-allowed TRS breaking channels with similar predicted thermodynamic consequences~\cite{Annett1990}. Thus, in general it can be very hard to pin down the exact symmetry of the superconducting order parameter by routine probes of unconventional SCs in a real material.

\subsection{Example of a TRS breaking state: a chiral superconductor}

We now illustrate how to use the GL formalism outlined above to deduce the possible TRS breaking states of a SC through a particular example. 

Chiral SCs~\cite{Kallin2016} are defined as SCs for which the phase of the order parameter winds in a certain direction (clockwise or anticlockwise) while moving in the $\bk$-space about some axis on the Fermi surface of the underlying metal. The case of a TRS breaking state resulting from a multi-dimensional irrep of the point group is thus very interesting, since it necessarily gives rise to a type of superconducting ground state which is chiral in nature. Also, in this case, since the superconducting ground state breaks additional crystal symmetries, it in general needs an unconventional pairing mechanism (not electron-phonon).

We will now discuss the specific example of constructing TRS breaking superconducting channels in a tetragonal system with the point group  symmetry $D_{4h}$. We will consider the simplest case of symmorphic space group and strong SOC in the following example which is probably relevant for the case of Sr$_2$RuO$_4$~\cite{Mackenzie2003}. The D$_{4h}$ point group has $8$ one dimensional irreps (4 of them have even parity and the other 4 have odd parity) and 2 two dimensional irreps (one with even parity denoted by E$_g$ and the other with odd parity denoted by E$_u$). Centrosymmetry implies that the possible superconducting states are either purely triplet or purely singlet states.  Furthermore, a TRS breaking superconducting order parameter is thus possible only in the E$_g$ or the E$_u$ irrep. We will now focus only on these two irreps and construct the possible TRS breaking superconducting order parameters for the system. 

\begin{table}
\centering
\caption{Important basis functions of D$_{4h}$ for the 2 two dimensional irreps corresponding to strong SOC. $A_1$, $A_2$ and $B_2$ are constants independent of $\bk$.\label{tab:D4h_basis}} 
\begin{tabular}{|c||c|c|}
\hline
D$_{2d}$ & \multicolumn{2}{c|}{Basis functions} \\
\hline
Irreps & Scalar (even) & Vector (odd) \\\hline\hline
$E_g$ & $A_1 k_z \left(\begin{array}{c} k_x \\ k_y\end{array}\right)$ & -- \\\hline
$E_u$ & -- & $\left(\begin{array}{c} A_2 k_z \hat{x} + B_2 k_x \hat{z} \\ A_2 k_z \hat{y} + B_2 k_y \hat{z}\end{array}\right)$ \\\hline
\end{tabular}
\end{table}

\begin{figure}[t]
\centerline{
\includegraphics[width=0.30\textwidth]{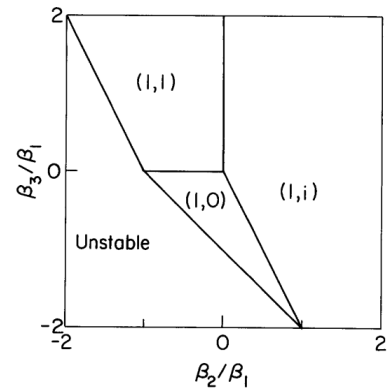}}
\caption{(Color online) The phase diagram corresponding to the two dimensional irreps of $D_{4h}$, reproduced with permission from  Ref.~\onlinecite{Annett1990}.}
\mylabel{fig:phase_diagram}
\end{figure}

\begin{figure*}[t]
\centerline{
\includegraphics[width=0.99\textwidth]{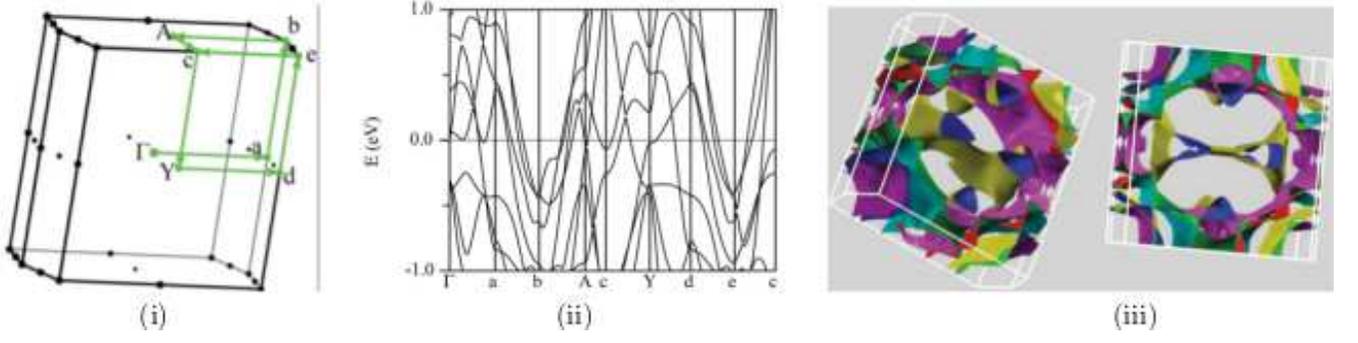}}
\caption{\mylabel{fig:LaNiGa2}(Color online) Band structure of LaNiGa$_2$. The panels (i), (ii) and (iii) show the unit cell, band structure near the Fermi level and the corresponding Fermi surfaces without spin-orbit coupling. These plots are reproduced with permission from Ref.~\onlinecite{Singh2012}. Copyright 2012 by the American Physical Society.}
\end{figure*}

We consider strong SOC and then the possible even scalar and odd vector basis functions for the E$_g$ and E$_u$ irreps are listed in \tab{tab:D4h_basis}. The fourth order invariant corresponding to the 2 two dimensional irreps of $D_{4h}$ gives the quartic order term of the GL free energy~\cite{Annett1990,Sigrist1991} to be
\beq\mylabel{eq:fenergy}
f_4 = \beta_1 (|\eta_1|^2 + |\eta_2|^2)^2 + \beta_2 |\eta^2_1 + \eta^2_2|^2 + \beta_3 (|\eta_1|^4 + |\eta_2|^4)
\eeq
where $(\eta_1, \eta_2)$ are the two complex components of the two dimensional order parameters. This free energy now needs to be minimized with respect to both $\eta_1$ and $\eta_2$. To this end, we write $\eta_j = |\eta_j|e^{i \alpha_j}$ where $|\eta_j|$ is the real amplitude and $\alpha_j$ is the phase ($j = 1$ and $2$). Then we define $|\eta_1| = |\eta|\cos(\theta)$, $|\eta_2| = |\eta|\sin(\theta)$ and $\alpha = (\alpha_1 - \alpha_2)$. Thus $\theta$ determines the relative amplitude and $\alpha$ determines the relative phase of the two components and we take $0 \leq \theta \leq \pi/2$ and $0 \leq \alpha \leq 2\pi$. Then we have 
\beq
(\eta_1,\eta_2) = |\eta| e^{i\alpha_1} [\cos(\theta), \sin(\theta)e^{i\alpha}].
\eeq
Thus effectively we need to minimize the free energy with respect to the two real variables $\theta$ and $\alpha$. Using the above equation in \eqn{eq:fenergy}, the free energy simplifies to 
\beq
f_4 = |\eta|^4 \left[ (\beta_1 + \beta_2 + \beta_3) -\frac{1}{2} \sin^2(2\theta) \{\beta_3 - 2\beta_2 \sin^2(\alpha)\}  \right].
\eeq
To minimize the above free energy with respect to $\theta$ and $\alpha$, we compute and equate the corresponding first derivatives to zero which gives
\bea
\frac{\partial f_4}{\partial \theta} &=& 0 \non\\
\Rightarrow \sin(4\theta) \{ \beta_3 - 2\beta_2 \sin^2(\alpha)\} &=& 0
\eea 
and 
\bea
\frac{\partial f_4}{\partial \alpha} &=& 0 \non\\
\Rightarrow \beta_2 \sin^2(2\theta)\sin(2\alpha)\} &=& 0
\eea 
assuming $|\eta|\neq0$. The above two equations give the following nontrivial and inequivalent solutions: $(\eta_1,\eta_2) = (1,0)$, $\frac{1}{\sqrt{2}}(1,1)$ and $\frac{1}{\sqrt{2}}(1,i)$. The corresponding phase diagram is shown in \fig{fig:phase_diagram} (reproduced from Ref.~\onlinecite{Annett1990}). We note that there is an extended region in the parameter space where the states corresponding to $(\eta_1,\eta_2) = \frac{1}{\sqrt{2}} (1,i)$ is stabilized. The instabilities corresponding to this case spontaneously break TRS at $T_c$ due to a nontrivial phase difference between the two order parameter components. The even parity TRS breaking superconducting order parameter belonging to $E_g$ is given by
\beq\mylabel{eqn:singlet}
\Delta(\bk) = \Delta_0 k_z (k_x + i k_y)
\eeq
where $\Delta_0$ is the real amplitude independent of $\bk$. This is a \emph{chiral d-wave} singlet order parameter. The odd parity superconducting order parameter belonging to $E_u$ gives rise to the TRS breaking triplet state with the $d$-vector given by 
\beq\mylabel{eqn:triplet}
\bd(\bk) = \left[ A k_z, i A k_z, B(k_x + i k_y)\right].
\eeq 
Here, $A$ and $B$ are material dependent real constants independent of $\bk$ and in general they are nonzero. We note that the values of $A$ and $B$ determine the orientation of the $d$-vector. For example, for $A=0$ the d-vector points along the $c$-axis and for $B=0$ the $d$-vector points in the $ab$-plane. We also note that
\beq
\bd(\bk) \times \bd^*(\bk) = 2 i A k_z (B k_x \hat{x} - B k_y \hat{y} - A k_z \hat{z})
\eeq 
which is nonzero in general. Hence, this superconducting state is a \emph{nonunitary chiral p-wave} triplet state. The special case of $A=0$ is the unitary chiral p-wave triplet case, which was previously applied to the case of Sr$_2$RuO$_4$ (see Sec.~\ref{CorrSec} for an overview of recent developments).

\subsection{Band structure}
Here, we discuss the insights which can be gained about the pairing symmetry of a TRS breaking SC from band structure calculations using first principles density functional theory (DFT). Band structures of several TRS breaking SCs have been reported, examples include LaNiGa$_2$~\cite{Singh2012}, Re$_6$Zr~\cite{Khan2016} and La$_7$Ir$_3$~\cite{Li2018}. We discuss the case of LaNiGa$_2$ shown in the \fig{fig:LaNiGa2} (reproduced from  Ref.~\onlinecite{Singh2012}) as an example.

A common feature of most of these materials is that there are multiple bands crossing the Fermi level leading to several Fermi surface sheets. These bands can arise from several orbitals of one or more types of atoms in the several symmetry related sites. The information about the most important orbitals, for example, can be obtained by computing the projected density of states (DOS) of these orbitals and looking at their contribution to the total density of states at the Fermi level. For the case of LaNiGa$_2$, there are several bands crossing the Fermi level as seen from the \fig{fig:LaNiGa2}(ii) and they arise mainly from the $p$-orbitals of Ga and $d$-orbitals of Ni. As a result the shape of the Fermi surface sheets are quite complicated in this example as shown in  \fig{fig:LaNiGa2}(iii).

The topology of the Fermi surfaces plays an important role in determining the thermodynamic properties of the TRS breaking superconducting ground states. For example, even if an order parameter features nodes at certain regions of the Brillouin zone if the Fermi surfaces are open or have necks in that region, the thermodynamic features of the system will behave similar to that of a fully gapped state~\cite{Bhattacharyya2015}. 

While analyzing the features of the band structure near the Fermi level, it is also important to take note of the effects of SOC. If SOC produces significant splitting of the bands near the Fermi level which can change the topology of the Fermi surfaces, for example, then it is imperative to consider the effect of SOC in the pairing symmetry of the order parameter. And, as we discussed earlier, SOC can have dramatic effects on the pairing symmetry depending on the presence or absence of inversion symmetry in the material.

DFT calculations can also be used to fit the band structure near the Fermi level to a tight-binding Hamiltonian including a ``few'' most important orbitals of the material. Then for a ``manageable'' size of the Hamiltonian one can make sure that the normal state band structure near the Fermi level, the topology of the Fermi surfaces and the density of states near the Fermi level is faithfully represented by this normal state Hamiltonian. This normal state Hamiltonian can then be used in conjunction with the pairing models deduced from symmetry to predict and compare experimentally observable quantities.

\subsection{Estimation of internal magnetic field/magnetic moment}

Estimating the strength of the internal magnetic field or the magnetic moment is in general a nontrivial task. This is because of the following reasons.
\begin{enumerate}
\item \label{mu1} Firstly, the magnetic moment per unit cell of the superconducting state $\mu_s$ depends on the details of the pairing model. For example, for a chiral SC such as one with the chiral $p$-wave triplet state, the magnetic moment can be estimated from the corresponding orbital angular momentum~\cite{Miyake2017,Robbins2018}. For a material with the loop-super current ground state~\cite{Ghosh2018} proposed to be realized in Re$_6$(Zr, Hf, Ti), the spontaneous internal magnetic field can be estimated from the super-current in the ground state. For LaNiGa$_2$, from a semi-first principles approach we can directly compute the spontaneous magnetic moment due to the non-unitary triplet state~\cite{Ghosh2019}.
 
\item \label{mu2} Secondly, even if the magnetic moment $\mu_s$ can be accurately predicted, the muon does not measure $\mu_s$ but the induced internal field $\mathbf{B}_{\rm int}(\mathbf{r})$ which depends, on an atomic scale, on the location $\mathbf{r}$ of the muon within the unit cell-- an averaged magnetic moment is therefore not enough to make a quantitative prediction. 

\item \label{mu3} Thirdly, given that the muon is a strong local perturbation, an understanding of the way the muon changes the local crystal and electronic structure of the sample is essential for a quantitative prediction. 
\end{enumerate}

While a naive estimate of the internal field $B_{\rm int} \sim \mu_0 \mu_s / 4\pi abc $ gives the right order of magnitude for Re$_6$Zr~\cite{Ghosh2018} and  LaNiGa$_2$~\cite{Ghosh2019} overcoming the issues in points \ref{mu2} and \ref{mu3}, above, is essential in order to develop a quantitative understanding of zero-field $\mu$SR in SCs with broken TRS. Indeed, in SCs any intrinsic fields ought to be, in the bulk, fully screened by the Meissner field, so the only reasonable interpretation of the ability of the muon to detect the intrinsic field is that the muon itself locally suppresses the order parameter. In this picture, what the muon sees in fact is the screening field, rather than the intrinsic magnetisation~\cite{Miyake2018}. Ref.~\cite{Miyake2017} contains the only quantitative calculation realizing this scenario, in a very simplified model of a chiral SC. Meanwhile current DFT studies of the effect of the muon on the local crystal and electronic structure~\cite{Foronda2015,Onuorah2019} have not yet been extended to SCs.

\section{mean field theory}
\mylabel{sec:MFT}
In this section we review the generalized BCS/Bogoliubov-de Gennes theory~\cite{Sigrist1991}, which is an essential tool for computing the excitation spectra and predicting experimental  observables for a particular superconducting ground state. 

\subsection{General formalism}

A general Hamiltonian describing pairing of electrons in momentum space can be written as
\bea
\mathcal{H} &=& \sum_{\bk,\sigma} \xi(\bk) c^\dagger_{\bk,\sigma}c_{\bk,\sigma} \\ 
&+& \frac{1}{2}\sum_{\bk,\bk',\bar{\bq},\{\sigma_i\}}  V_{\{\sigma_i\}}(\bk,\bk') c^\dagger_{\frac{\bar{\bq}}{2}+\bk,\sigma_1} c^\dagger_{\frac{\bar{\bq}}{2}-\bk,\sigma_2} c_{\frac{\bar{\bq}}{2}-\bk',\sigma_3}c_{\frac{\bar{\bq}}{2}+\bk',\sigma_4}\non
\eea
where $c^\dagger_{\bk,\sigma}$ creates an electron with momentum $\bk$ and spin $\sigma$ and $\xi(\bk) = (\epsilon(k) - \mu)$ with $\mu$ being the chemical potential and $\epsilon(\bk)$ being the normal state dispersion. $V_{\sigma_1,\sigma_2,\sigma_3,\sigma_4}(\bk,\bk')$ describes the elements of the pairing interaction $\hat{V}$ between two electrons and includes the possibility of formation of pairs with finite total momentum $\bar{\bq}$. We will, however, consider only the case of zero total momentum ($\bar{\bq}=0$) pairs with the only nonzero matrix elements
\beq
V_{\sigma_1,\sigma_2,\sigma_3,\sigma_4}(\bk,\bk') = \left<\bk,\sigma_1;-\bk,\sigma_2|\hat{V}|-\bk',\sigma_3;\bk',\sigma_4\right>.
\eeq
The pairing interaction $\hat{V}$ can have different origins including being mediated by phonons~\cite{Tinkham1996} or spin-fluctuations~\cite{Scalapino2012} for example. The important point to note is that the interaction is attractive within a narrow window near the Fermi level and the Fermi surface is unstable to this attractive interaction leading to the formation of Cooper pairs. To find the ground state properties of the system, we decompose the quartic interaction term by introducing mean-fields using a standard Curie-Weiss type mean-field theory~\cite{Tinkham1996} and ignoring fluctuations above the mean-field ground state ---although fluctuations above the mean-field ground state can be systematically taken into account when necessary~\cite{Coleman2015}. We define the mean-field pairing potential or the gap function as
\beq\mylabel{eqn:gap_eqn}
\Delta_{\sigma,\sigma'}(\bk) = \sum_{\bk',\sigma_1,\sigma_2} V_{\sigma,\sigma',\sigma_1,\sigma_2}(\bk,\bk')\left< c_{-\bk',\sigma_1} c_{\bk',\sigma_2}\right>
\eeq
with $\Delta_{\sigma,\sigma'}(\bk)$ being the elements of the gap matrix $\hat{\Delta}(\bk)$ satisfying \eqn{gap_antisymmtery}. Introducing these mean-fields, the Hamiltonian within the mean-field approximation takes the form
\beq
\mathcal{H}_{MF} = \sum_{\bk,\sigma} \xi(\bk) c^\dagger_{\bk,\sigma}c_{\bk,\sigma} + \half \sum_{\bk,\sigma,\sigma'}[\Delta_{\sigma,\sigma'}(\bk)c^\dagger_{\bk,\sigma} c^\dagger_{-\bk,\sigma'} + h. c. ].
\eeq
We now use the Bogoliubov-de Gennes (BdG) formalism to find the ground state of this mean-field Hamiltonian. To this end, we define the Nambu spinor as
\beq
\Psi_{\bk} =  \begin{pmatrix}\hat{c}_{\bk} \\ \\\hat{c}^{\dagger T}_{-\bk}\end{pmatrix} \,\,\,\text{with} \,\, \hat{c}_{\bk} = \begin{pmatrix} c_{\bk\uparrow} \\\\c_{\bk\downarrow} \end{pmatrix}.
\eeq 
Then the mean-field Hamiltonian takes the form
\beq\mylabel{eqn:HMF_Nambu}
\mathcal{H}_{MF} = \half \sum_{\bk} \Psi^\dagger_{\bk} \mathcal{H}_{BdG}(\bk) \Psi_{\bk} + E'_g
\eeq
where $E'_g = \sum_{\bk}\xi(\bk)$ gives a constant shift of the ground state energy and will be ignored from here on. Defining $\hat{\xi}_{\bk} = \xi(\bk) \mathcal{I}_{2}$ with $\mathcal{I}_n$ being the identity matrix of order $n$, the BdG Hamiltonian can be written as
\beq
\mathcal{H}_{BdG}(\bk) = \begin{pmatrix} \hat{\xi}_{\bk} & \hat{\Delta}(\bk) \\ \\ \hat{\Delta}^\dagger(\bk) & -\hat{\xi}_{\bk} \end{pmatrix}.
\eeq
This Hamiltonian can be diagonalized using a unitary transformation defined as
\bea
\hat{U}^{\dagger}_{\bk} \mathcal{H}_{BdG}(\bk) \hat{U}_{\bk} &=& \hat{\mathcal{E}}_{\bk} \,\,;\,\,\,\hat{U}^{\dagger}_{\bk}\hat{U}_{\bk} = \hat{U}_{\bk} \hat{U}^{\dagger}_{\bk} = \mathcal{I}_{4}\\\non\\
\text{with}\,\,\,\hat{\mathcal{E}}_{\bk} &=& \begin{pmatrix} E_{\bk,1}&0&0&0 \\ 0&E_{\bk,2}&0&0 \\ 0& 0& -E_{\bk,1} &0 \\ 0& 0& 0& -E_{\bk,2}\end{pmatrix}.
\eea
$E_{\bk,\alpha}$ are dispersions of the two branches of the Bogoliubov quasiparticles labeled by $\alpha = 1$ and $2$ given by
\beq\mylabel{eqn:quasi_dispersion}
E_{\bk,\alpha} = \sqrt{\xi^2(\bk) + \half Tr[\Delta(\bk) \Delta^{\dagger}(\bk)] + (-1)^{\alpha} |\bq(\bk)|}.
\eeq
The mean-field Hamiltonian in \eqn{eqn:HMF_Nambu} can then be written as
\beq\mylabel{eqn:HMF_bogo}
\mathcal{H}_{MF} = \half \sum_{\bk} \Gamma^\dagger_{\bk} \hat{\mathcal{E}}_{\bk} \Gamma_{\bk} = \sum_{\bk,\alpha} E_{\bk,\alpha} \gamma^{\dagger}_{\bk,\alpha} \gamma_{\bk,\alpha} + E''_g
\eeq
where we have defined the creation operator of the Bogoliubov quasiparticles with momentum $\bk$ in branch $\alpha$ as $\gamma^{\dagger}_{\bk,\alpha}$ and 
\beq
\Gamma_{\bk} = \hat{U}^{\dagger}_{\bk} \Psi_{\bk} = \begin{pmatrix}\hat{\gamma}_{\bk} \\ \\\hat{\gamma}^{\dagger T}_{-\bk}\end{pmatrix} \,\,\,\text{with} \,\, \hat{\gamma}_{\bk} = \begin{pmatrix} {\gamma}_{\bk,1} \\\\{\gamma}_{\bk,2} \end{pmatrix}.
\eeq
$E''_g = \half\sum_{\bk,\alpha} E_{\bk,\alpha}$ also gives a constant shift in the ground state energy and will be ignored in the following discussion. The Bogoliubov quasiparticle operators obey fermionic anti-commutation relations
\beq
\left\{ \gamma^{\dagger}_{{\bk}_1,\alpha_1}, \gamma_{{\bk}_2,\alpha_2}\right\} = \delta_{{\bk}_1,{\bk}_2} \delta_{{\alpha}_1,{\alpha}_2}.
\eeq
We note from \eqn{eqn:HMF_bogo} that the mean-field ground state is a free Fermi gas of Bogoliubov quasiparticles and we can define the thermal averages in this ground state as
\beq
\left< \gamma^{\dagger}_{{\bk}_1,\alpha_1} \gamma_{{\bk}_2,\alpha_2}\right> = \delta_{{\bk}_1,{\bk}_2} \delta_{{\alpha}_1,{\alpha}_2} \mathit{f}(E_{{\bk}_1,\alpha_1})
\eeq
where $\mathit{f}(x) = (1+e^{\beta x})^{-1}$ is the Fermi-Dirac distribution function with $\beta = (k_B T)^{-1}$. Taking the form
\beq
\hat{U}_{\bk} = \begin{pmatrix}\hat{u}_{\bk} & \hat{v}_{\bk} \\ \hat{v}^*_{-\bk} & \hat{u}^*_{-\bk} \end{pmatrix},
\eeq
we can explicitly write the Bogoliubov transformation of the fermion operators as
\beq
c_{\bk,\sigma} = \sum_{\sigma'} \left[ u_{\sigma,\sigma'}(\bk) \gamma_{\bk,\sigma'} + v_{\sigma,\sigma'}(\bk) \gamma^{\dagger}_{-\bk,\sigma'} \right].
\eeq
This expression of the fermion operators can now be used to find the expectation value of any operator in the mean-field ground state. In particular, \eqn{eqn:gap_eqn} leads to the self-consistency equation for the gap function and determines the temperature dependence of $\hat{\Delta}(\bk)$ and $T_c$.

We note from \eqn{eqn:quasi_dispersion} that the two branches of the quasiparticle dispersions are degenerate for singlet and unitary-triplet pairings but a non-unitary triplet pairing state lifts this degeneracy leading to two momentum dependent gaps $\sqrt{\half Tr[\Delta(\bk) \Delta^{\dagger}(\bk)] \pm |\bq(\bk)|}$. This splitting is a consequence of reduction in the symmetry in the superconducting state due to broken TRS caused by the non-unitary pairing.

\subsection{Low-temperature thermodynamics characterizing unconventional superconductivity}
To determine the behaviors of the experimental observables, such as specific heat ($C$), penetration depth ($\lambda$), NMR relaxation rate ($1/T_1$) and superfluid density ($\rho_s$), which are routinely measured to characterize unconventional SCs, it is important to consider the behavior of the DOS of the quasiparticle excitations (qpDOS)~\cite{Sigrist1991}. Depending on the symmetry of the superconducting order parameter, the low energy behavior of the qpDOS can be qualitatively different and thus leads to characteristic low temperature behaviors of the experimental observables. The qpDOS is defined as
\beq
g(E) = \sum_{\bk,\alpha} \delta(E-E_{\bk,\alpha}).
\eeq

First, let us discuss the case of conventional rotationally symmetric $s$-wave SCs. In this case, the system is fully gapped and only with energy above the energy gap excitations are possible. As a result, the qpDOS in this case is given by
\beq
g(E) = N(0) \frac{E}{\sqrt{E^2 - \Delta(0)^2}} \Theta(E-\Delta(0))
\eeq
where $N(0)$ is the normal state DOS at the Fermi level and $\Delta(0)$ is the size of the gap at zero temperature. Hence, the qpDOS is zero within the gap and has a coherence peak at $E=\Delta(0)$ visible in single particle tunneling experiments~\cite{Coleman2015} for example. In general, the above type of qpDOS leads to an exponential suppression in the observables at low temperatures.

For an unconventional SCs, however, the presence of nodes in the quasiparticle spectrum leads to a starkly different qpDOS. The low energy behavior~\cite{Sigrist1991,Fernandes2011,Mazidian2013} of the qpDOS for an unconventional SC is
\bea
g(E) &\propto& E \,\,\,\,\text{for line nodes,}\mylabel{eqn:line_DOS}\\
&\propto& E^2 \,\,\,\,\text{for point nodes.}\mylabel{eqn:point_DOS}
\eea
This type of low energy qpDOS in general leads to a power law dependence in the low temperature properties of the observables. The low temperature dependence of the observables under consideration can be computed from the following relations~\cite{Fernandes2011} valid for $T \ll T_c$:
\bea
C(T) &\propto& \beta \int_0^\infty dE\,\,g(E) E^2 \left(-\frac{\partial f(E)}{\partial E}\right), \\
\Delta\lambda(T) &\propto& \int_0^\infty dE\,\,g(E) \left(-\frac{\partial f(E)}{\partial E}\right), \\
\frac{1}{T_1 T} &\propto& \int_0^\infty dE\,\,g^2(E) \left(-\frac{\partial f(E)}{\partial E}\right), \\
\bigg(\frac{\rho_s(T)}{\rho_s(0)} &-& 1 \bigg) \propto -\Delta\lambda(T) 
\eea
where $\Delta\lambda(T) = [\lambda(T) - \lambda(0)]$. Using the expressions of the qpDOS from \eqn{eqn:line_DOS} and \eqn{eqn:point_DOS} in the above expressions we can now compute the low temperature behaviors of the observables~\cite{Sigrist1991,Fernandes2011,Mazidian2013}:
\bea
C(T) &\propto& T^2 \,\,\,\,\text{for line nodes,}\mylabel{Low-T-C-lines}\\
&\propto& T^3 \,\,\,\,\text{for point nodes.}\mylabel{Low-T-C-points}\\
\Delta\lambda(T) &\propto& T \,\,\,\,\text{for line nodes,}\mylabel{Low-T-Lambda-lines}\\
&\propto& T^2 \,\,\,\,\text{for point nodes.}\mylabel{Low-T-Lambda-points}\\
\frac{1}{T_1 T} &\propto& T^2 \,\,\,\,\text{for line nodes,}\mylabel{Low-T-T1-lines}\\
&\propto& T^4 \,\,\,\,\text{for point nodes.}\mylabel{Low-T-T1-points}
\eea

As noted above the full mean-field description of a given system requires the self-consistent solution of the BdG equations and the mean field self-consistency equations. Nevertheless we note that a lot of information can be obtained by solving the BdG equations for a given form of the pairing potential, obtained by symmetry analysis. Unlike the self-consistency equations, the BdG equations do not feature the pairing interaction. Indeed the results in  Eqs.~(\ref{Low-T-C-lines}, \ref{Low-T-C-points}, \ref{Low-T-Lambda-lines}, \ref{Low-T-Lambda-points}, \ref{Low-T-T1-lines} and \ref{Low-T-T1-points}) depend only on the basic features of the pairing potential and Fermi surface topology (shape), and therefore can be used to correlate experimental data with a GL analysis without recourse to a model of the electron-electron pairing interaction. This is an extremely useful weapon in the arsenal of the theorist trying to pry information about the pairing state of a SC out of the available experimental data. 

\section{Novel ground states in multiband superconductors}
\mylabel{sec:novel-gs}

Superconductivity in systems with multiple bands is usually considered to be effectively occurring in a single band with a gap which has nontrivial momentum dependence in the different Fermi surface sheets of the material as discussed in the previous sections. But recent studies~\citep{Nomoto2016,Brydon2016,Yanase2016,Nica2017,Agterberg2017,Brydon2018,Huang2019,Ramires2019} in many multiband systems such as the iron-based SCs, half-Heusler compounds, UPt$_3$ and Sr$_2$RuO$_4$ have pointed out that internal electronic degrees of freedom, coming from the orbitals or sublattice for example, can give rise to a nontrivial structure to the Cooper pair wave function and in particular to Cooper pairs with higher spins. However, this type of nontrivial superconducting pairing state is usually hard to distinguish from its trivial counterpart since both have qualitatively similar low energy properties.

\subsection{Internally-antisymmetric nonunitary triplet pairing in LaNiC$_2$ and LaNiGa$_2$}
We emphasize two cases, LaNiC$_2$ and LaNiGa$_2$, where the group-theoretical analysis seems to be at odds with the experimental results for TRS breaking superconducting order parameters. In both the cases, the symmetry analysis suggests a nodal TRS-breaking nonunitary triplet superconducting state~\cite{Hillier2009,Quintanilla2010,Hillier2012} while experiments suggest a nodeless, two-gap behavior~\cite{Chen2013,Weng2016}.

\begin{figure}[t]
\centerline{
\includegraphics[width=0.25\textwidth]{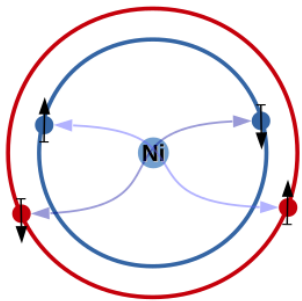}}
\caption{(Color online) Schematic of a non-unitary triplet pairing between same spins between two different Ni orbitals. Reproduced with permission from Ref. \onlinecite{Csire2018}.}
\mylabel{fig:ESP}
\end{figure}

 LaNiC$_2$ crystallizes in a noncentrosymmetric orthorhombic crystal structure~\cite{Hillier2009} with symmorphic space group $Amm2$ (No. $38$) and point group $C_{2v}$ whereas LaNiGa$_2$ crystallizes in a centrosymmetric orthorhombic crystal structure~\cite{Hillier2012} with symmorphic space group $Cmmm$ (No. $65$) and point group $D_{2h}$. Both the point groups $C_{2v}$ and $D_{2h}$ have only one dimensional irreps. Thus from the point group symmetry alone there is no degenerate instability channel which can potentially lead to a TRS breaking order parameter. Based on this observation, it was proposed~\cite{Hillier2009,Quintanilla2010,Hillier2012,Weng2016} that considering the effect of SOC to be negligible in both the materials, the three dimensional spin rotation group $SO(3)$ can provide a nonunitary triplet channel which by its very nature breaks TRS spontaneously at $T_c$. Based on the GL analysis~\cite{Hillier2012} it was shown that a paramagnetic coupling to the magnetisation can stabilize such a nonunitary triplet ground state. This nonunitary state in turn produces a pseudomagnetic field which splits the up and down spin Fermi surfaces generating an imbalance between the two spin species~\cite{Hillier2012,Miyake2014}. As a result a subdominant order parameter, induced magnetization, arises and it increases linearly near $T_c$ with decreasing temperature. But the corresponding nonunitary gap functions have nodes. This prompted the suggestion~\cite{Weng2016} that the pairing state may involve an orbital or band index. Such internal degrees of freedom \emph{cannot} be used to increase the dimensionality of the order parameter i.e. we still need a nonunitary triplet pairing state to break TRS-- which in any case is strongly supported by the observation in LaNiC$_2$ of the magnetization signal predicted~\cite{Hillier2012} by the GL theory of the nonunitary state. However, they can dramatically alter the quasi-particle spectrum by ensuring fermionic anti-symmetry without the need for a $\mathbf{k}$-dependent gap function, potentially leading to the required $\mathbf{k}$-independent gap. The pairing in the superconducting ground state can be between same spins but involving two different orbitals as shown for example with two $d$-orbitals of the Ni atom in the \fig{fig:ESP}. Thus the Cooper pair wave function is triplet in the spin sector but has isotropic even parity gap symmetry. Then the two gaps correspond to the two different spin species of different population rather than two different orbital character bands in this internally-antisymmetric nonunitary triplet (INT) ground state. A semi-phenomenological model of LaNiGa$_2$ featuring a detailed, {\it ab initio} description of the band structure and a \emph{single} adjustable parameter (namely the strength of the effective electron-electron, which is fixed by $T_c$) in the INT state reproduces the specific heat dependence on temperature very accurately~\cite{Ghosh2019} strongly suggesting that this model is an accurate description of the physics of this material.  
 
 The mechanism behind such an unusual pairing state remains to be determined, but measurements of LaNiC$_2$ under pressure suggested that the  superconductivity is situated on a dome, in close proximity to a correlated phase and quantum criticality \cite{Landaeta2017}. Interestingly, $\mu$SR measurements revealed evidence for nodal superconductivity in isostructural ThCoC$_2$, where it was suggested that the strong Fermi surface nesting favors an unconventional magnetically mediated pairing mechanism \cite{Bhattacharyya2019}. However, time reversal symmetry was found to be preserved in the superconducting state.

\begin{figure}[t]
\centerline{
\includegraphics[width=0.48\textwidth]{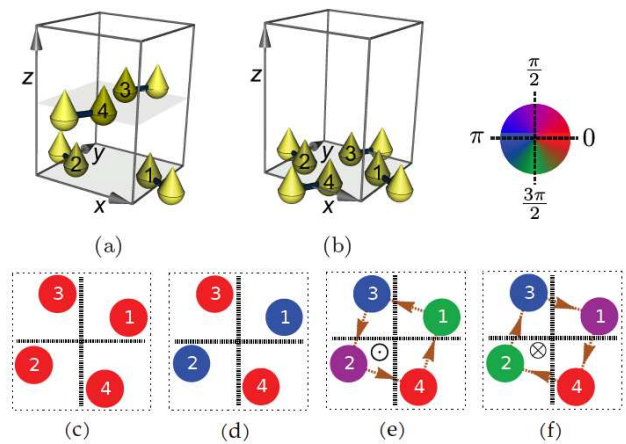}}
\caption{(Color online) Possible superconducting instabilities for a toy model (nonsymmorphic version (a) and symmorphic version (b)) with a tetragonal unit cell allowing for onsite values of the superconducting order parameter to be different in the four symmetrically distinct sites. (c)--(f): top view of the four symmetry-allowed superconducting instabilities for both models with the color wheel depicting the phase of the superconducting order parameter. The TRS-breaking instability is a linear combination of (e) and (f), which are degenerate. The arrows show the direction of the circulating loop super-currents within the unit cell. This figure is reproduced from Ref.~\onlinecite{Ghosh2018}.}
\mylabel{fig:LSC}
\end{figure}

\subsection{Loop-super current state}
This proposal~\cite{Ghosh2018} is based on a few key observations that the relevant materials (Re$_6$(Zr, Hf, Ti)~\cite{Singh2014,Singh2017,Singh2018}, Re$_{0.82}$Nb$_{0.18}$~\cite{ShangPRL2018} and La$_7$(Ir, Rh)$_3$~\cite{Barker2015,Li2018,Singh2018}) have several common features: several symmetry related sites (more than two is necessary) within the unit cell, several Fermi surfaces coming from several orbitals and a single fully gapped conventional BCS type thermodynamic behavior of observables. We consider a uniform onsite singlet pairing and allow for the possibility that the order parameter can potentially have different values (amplitude and phase) at the different symmetry related sites within the unit cell. Then using symmetry it is noted that there is a symmetry allowed ground state which has finite microscopic Josephson currents flowing between the symmetry related sites and is thus termed as a loop-super current (LSC) state. Such a ground state has inherent chirality and thus breaks TRS spontaneously at $T_c$. An example of such a symmetry allowed order parameter for a system with a $\mathcal{C}_4$ point group and four symmetrically distinct sites is shown in the \fig{fig:LSC}. Note that the degeneracy in one of the irrep of $\mathcal{C}_4$ in this case actually comes from TRS in the normal state itself. On symmetry grounds, such a ground state can be stabilized in Re$_6$(Zr, Hf, Ti)~\cite{Singh2014,Singh2017,Singh2018} and Re$_{0.82}$Nb$_{0.18}$~\cite{ShangPRL2018} but is not favorable energetically for La$_7$(Ir, Rh)$_3$~\cite{Barker2015,Li2018,Singh2018}. One of the attractions of the LSC state is that since it only relies on on-site pairing, it does not require an unconvenitonal pairing mechanism. On the other hand, the question of competition with other, more conventional states cannot be addressed within GL theory and is currently unexplored.

\subsection{Symmetry analysis including orbitals -- the case of Sr$_2$RuO$_4$}
In addition to the novel ground states proposed in the previous section, for some materials in particular Sr$_2$RuO$_4$ different experiments with high quality samples give apparently conflicting results~\cite{Mackenzie2017}. It has now become increasingly clear that the common chiral $p$-wave triplet state~\cite{Mackenzie2003} with $\bd$-vector pointing along the $c$-axis proposed in this material based on the usual symmetry classification discussed earlier can not systematically account for all/most of the experimental signatures in this material as noted in \sect{sec:materials}. Based on this observation, recent studies~\cite{Huang2019,Ramires2019,kaba2019} have proposed that in this case it is essential to generalize the symmetry classification of the order parameters to include the orbital degrees of freedom as well. This way novel symmetrically allowed possibilities of singlet, triplet with in-plane $\bd$-vector and mixed parity superconducting ground states~\cite{Huang2019,Ramires2019,kaba2019} have been proposed which may be able to explain all/most of the features in this material.

\begin{figure}[t]
\centerline{
\includegraphics[width=0.35\textwidth]{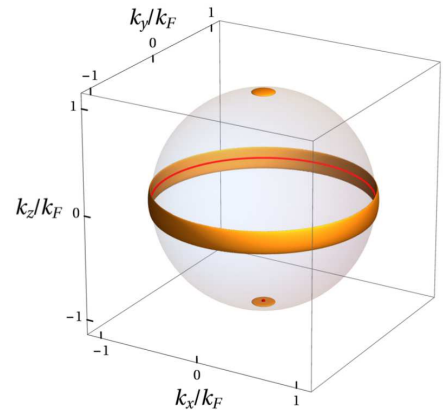}}
\caption{(Color online) Bogoliubov Fermi surfaces corresponding to the chiral d-wave singlet superconducting state given in \eqn{eqn:singlet}, shown here for the case where only one band has a spherical Fermi surface (semitransparent sphere) for a centrosymmetric multiband SC with generic symmetries and spontaneously breaking TRS at $T_c$. The point and line nodes (red dots and line, respectively), are ``inflated'' into spheroidal and toroidal Bogoliubov Fermi surfaces (orange surfaces) protected by a $Z_2$ topological invariant. Reproduced from  Ref.~\onlinecite{Agterberg2017}. Copyright 2017 by the American Physical Society.}
\mylabel{fig:BFS}
\end{figure}

\subsection{Bogoliubov Fermi surfaces}

It was shown recently~\cite{Agterberg2017,Brydon2018} that for a multiband SC with centrosymmetry and spontaneously broken TRS at $T_c$, considering the internal electronic degrees of freedom in Cooper pair formation in general leads to only two possible types of quasiparticle excitation gaps: either the system is fully gapped or it has Bogoliubov Fermi surfaces (BFS). These are two dimensional surfaces which are generated, for example, by inflating a point or a line node by a pseudo-magnetic field inherent to the TRS breaking nodal superconducting state. This pseudo-magnetic field however is proportional to the second order of interband pairing. As a result its strength is quite small. But, the BFSs are protected by a $Z_2$ topological invariant and hence cannot be removed by symmetry preserving perturbations. A superconducting state possessing BFSs can have characteristic experimental signatures~\cite{Lapp2019} such as a nonzero residual DOS at the Fermi level which can potentially be measured in a single particle tunneling experiment and unconventional thermodynamic signatures at low temperatures. But since the strength of the pseudo-magnetic field which basically distinguishes these type of states from their ``trivial'' counterparts is generically small, it may turn out to be difficult to distinguish them in practice.

\section{Summary and outlook}
\mylabel{sec:summary}
In this review we have discussed unconventional SCs which have been recently discovered to break time-reversal symmetry, in addition to gauge symmetry. The picture that emerges from these systems is complex and diverse, but at the same time there are some elements in common and that distinguish them from earlier examples of broken TRS in correlated SCs. First of all, the SCs in question appear, in many ways, conventional: they have nodeless quasiparticle spectra; $T_c$ is insensitive to non-magnetic impurities; and their normal state is often an ordinary Fermi liquid. Secondly, they often have multi-band electronic structures, often with complex unit cells having many symmetry-related atomic sites. These suggest that there are common mechanisms for broken TRS which present themselves in multi-orbital SCs. 

The small family formed by LaNiC$_2$ and LaNiGa$_2$ deserves a special mention. While it is usually very hard to pin down what the pairing symmetry is for a given unconventional SC, there has been remarkable progress in the last few years on these two compounds. These systems have very low symmetry, and as a result there are very few possibilities to choose from. The observation of broken TRS at $T_{\rm c}$ from zero-field $\mu$SR is only compatible with equal-spin, nonunitary triplet pairing. This led to the prediction of a sub-dominant magnetization accompanying the superconducting order parameter, which was subsequently confirmed experimentally. The nodeless, two-gap behaviour observed in these systems further points to a common origin of the superconductivity and the excellent agreement of a semi-phenomenological theory based on these ideas with a single adjustable parameter with measurements suggests that these are the first TRS-breaking SCs for which we have a complete and accurate picture of the superconducting state-- though understanding the origin of the pairing interaction remains a mystery. 
 
While many SCs with broken TRS show other signs of unconventional pairing, independent confirmation specifically of broken TRS is usually hard to obtain. It has been obtained through the optical Kerr effect in Sr$_2$RuO$_4$ and UPt$_3$ and PrOs$_4$Sb$_{12}$, as well as through bulk SQUID measurements of magnetization in LaNiC$_2$. Obtaining such evidence for other systems is therefore extremely urgent, as is a better understanding of the way the muon interacts with the crystal and with the superconducting condensate. In this respect we wish to highlight the challenge posed by the very consistent relation between the nuclear moments and muon spin relaxation rate in Re-based SCs~\cite{ShangPRL2018}.

We also note that many of these materials are in the dirty limit. Their resilience in the face of disorder distinguishes them quite clearly from other (particularly nodal) unconventional SCs. An understanding of the effect of disorder on the exotic states described in the last part of this review is a pending issue that will need to be addressed. 

To conclude, we note that the macroscopic quantum coherence of SCs makes them inherently useful materials out of which to build qubits~\cite{devoret2004superconducting}-- indeed, recently the historic milestone of quantum supremacy has recently been claimed using a superconducting quantum computer ~\cite{arute2019quantum}. Looking at the future, unconventional SCs offer additional degrees of freedom and therefore potentially could offer routes to novel qubit architectures. The varied phenomenology of the SCs with broken TRS reviewed here suggests that they may be fertile ground to essay such technologies, in particular in the cases where TRS breaking potentially occurs in a triplet pairing state. Realizing such promise will require not only detailed, quantitative theories of the equilibrium state of such systems, but also a precise understanding of their dynamics. For instance, one could investigate whether the net magnetization of a nonunitary triplet pairing SC like LaNiC$_2$ or LaNiGa$_2$ may show coherent oscillations (in contrast to the magnetization of a ferromagnet, which relaxes quickly). That could make possible qubits based on spin, rather than charge currents. Likewise the chirality of loop super-currents could potentially be used to encode quantum information as well.

\section{Acknowledgments}
We acknowledge  K. Miyake and F. Steglich for fruitful discussions. This work was supported by the National Key R$\&$D Program of China (Grants No.~2017YFA0303100 and No.~2016YFA0300202), and the National Natural Science Foundation of China (No.~11874320,  No.~11974306 and No.~U1632275). SKG, JQ and JFA are supported by EPSRC through the project ``Unconventional Superconductors: New paradigms for new materials'' (grant references EP/P00749X/1 and EP/P007392/1).

%

\end{document}